\documentclass[conference]{IEEEtran}
\IEEEoverridecommandlockouts
\usepackage{cite}
\usepackage{amsmath,amssymb,amsfonts}
\usepackage{algorithmic}
\usepackage{graphicx}
\usepackage{textcomp}
\usepackage{xcolor}
\def\BibTeX{{\rm B\kern-.05em{\sc i\kern-.025em b}\kern-.08em
    T\kern-.1667em\lower.7ex\hbox{E}\kern-.125emX}}
\usepackage{subcaption}
\usepackage{hyperref}

\usepackage[most]{tcolorbox}

\usepackage{enumitem}
\usepackage[normalem]{ulem}
\usepackage{adjustbox}
\usepackage{url}
\usepackage{yquant}
\usepackage{mhchem,multirow,calc,array,booktabs}
\newlength\mylenA
\newlength\mylenB
\newlength\mylenC

\usepackage[colorinlistoftodos,prependcaption,textsize=tiny]{todonotes} % TODO Remove

\usepackage{comment}

\makeatletter
\newcommand{\linebreakand}{%
  \end{@IEEEauthorhalign}
  \hfill\mbox{}\par
  \mbox{}\hfill\begin{@IEEEauthorhalign}
}
\makeatother

\newcommand\copyrighttext{%
  \footnotesize  This work is licensed to IEEE under the Creative Commons Attribution 4.0 (CC BY 4.0).}
\newcommand\copyrightnotice{%
\begin{tikzpicture}[remember picture,overlay]
\node[anchor=south,yshift=10pt] at (current page.south) {\fbox{\parbox{\dimexpr\textwidth-\fboxsep-\fboxrule\relax}{\copyrighttext}}};
\end{tikzpicture}%
}

\begin{document}

\DeclareRobustCommand{\hsout}[1]{\texorpdfstring{\sout{#1}}{#1}}
% First argument is the person suggesting the change (e.g. use your acronym)
\newcommand{\delete}[2]{\PackageWarning{Suggested change}{#1 wants to remove "#2"}\textcolor{red!80}{\hsout{#2}}}
\newcommand{\add}[2]{\PackageWarning{Suggested change}{#1 wants to add "#2"}\textcolor{green!70!black}{#2}}
\newcommand{\replace}[3]{\PackageWarning{Suggested change}{#1 wants to replace "#2" by "#3"}\textcolor{red!80}{\hsout{#2}}\textcolor{green!70!black}{#3}}

\tcbset{
  mytakeaway/.style={
    colback=gray!10,    % light grey background
    colframe=gray!70,   % darker grey border
    boxrule=0.5mm,      % border thickness
    arc=2mm,            % rounded corners
    left=1mm, right=1mm, top=1mm, bottom=1mm,
    fonttitle=\bfseries,
    title=Takeaway,
  }
}

\title{Evaluating the Effect of the Order of Optimization Passes in Quantum Circuit Optimization\\
}

% \author{
% \IEEEauthorblockN{1\textsuperscript{st} Xiao-Ting Michelle To}
% \IEEEauthorblockA{\textit{MNM-Team} \\
% \textit{Ludwig-Maximilians-Universität München}\\
% Munich, Germany \\
% michelle.to@nm.ifi.lmu.de}
% \and
% \IEEEauthorblockN{2\textsuperscript{nd} Nils Quetschlich}
% \IEEEauthorblockA{\textit{Chair for Design Automation} \\
% \textit{Technical University of Munich}\\
% Munich, Germany \\
% nils.quetschlich@tum.de}
% \and
% \IEEEauthorblockN{3\textsuperscript{rd} Amr Elsharkawy}
% \IEEEauthorblockA{\textit{Chair of Computer Architecture and Parallel Systems} \\
% \textit{Technical University of Munich}\\
% Garching bei München, Germany \\
% amr.elsharkawy@in.tum.de}
% \and
% \IEEEauthorblockN{4\textsuperscript{th} Robert Wille}
% \IEEEauthorblockA{\textit{Chair for Design Automation} \\
% \textit{Technical University of Munich}\\
% Munich, Germany \\
% robert.wille@tum.de}
% \and
% \IEEEauthorblockN{5\textsuperscript{th} Dieter Kranzlmüller}
% \IEEEauthorblockA{\textit{MNM-Team} \\
% \textit{Ludwig-Maximilians-Universität München}\\
% Munich, Germany \\
% kranzlmueller@ifi.lmu.de}
% }

\author{\IEEEauthorblockN{ Xiao-Ting Michelle To\IEEEauthorrefmark{1}, Nils Quetschlich\IEEEauthorrefmark{2}, Amr Elsharkawy\IEEEauthorrefmark{3}, Martin Schulz\IEEEauthorrefmark{3}\IEEEauthorrefmark{4}, Robert Wille\IEEEauthorrefmark{2} and Dieter Kranzlmüller\IEEEauthorrefmark{1}\IEEEauthorrefmark{4} }

\IEEEauthorblockA{\IEEEauthorrefmark{1}MNM Team, Ludwig-Maximilians-Universität in Munich, Munich, Germany\\
Email: michelle.to@nm.ifi.lmu.de, kranzlmueller@ifi.lmu.de}
\IEEEauthorblockA{\IEEEauthorrefmark{2}Chair for Design Automation, Technical University of Munich, Munich, Germany\\
Email: \{nils.quetschlich, robert.wille\}@tum.de}
\IEEEauthorblockA{\IEEEauthorrefmark{3}Chair of Computer Architecture and Parallel Systems, Technical University of Munich, Garching bei München, Germany\\
Email: \{amr.elsharkawy,schulzm\}@in.tum.de}
\IEEEauthorblockA{\IEEEauthorrefmark{4}Leibniz Supercomputing Centre, Garching bei München, Germany\\}}

\maketitle

\begin{abstract}
% This document is a model and instructions for \LaTeX.
% This and the IEEEtran.cls file define the components of your paper [title, text, heads, etc.]. *CRITICAL: Do Not Use Symbols, Special Characters, Footnotes, 
% or Math in Paper Title or Abstract.
% TODO
% Full paper: 8-10 pages; short paper: 4-6 pages

Quantum circuit optimization is critical for mitigating the noise inherent in current quantum hardware.
% reducing quantum circuit depth, single-qubit, and two-qubit gate count. 
Quantum compilers typically sequentially apply multiple optimizations (also called ``optimization passes'') to improve the circuit.
The impact of the order in which these passes are executed has yet been largely unexplored. 
This paper investigates the significance of the order of optimization passes within quantum circuit compilation, specifically analyzing interactions between different optimization methods and quantifying their mutual influences. 
Using Qiskit's compiler, we systematically evaluate pairwise combinations of 16 selected optimization passes, measuring circuit depth and gate count across various benchmark circuits. 
Our findings indicate dependencies between certain optimization passes, demonstrating that the order of the passes affects the optimization quality. 
In some cases, the worse performing sequence can be corrected through repeated pass application. %such that they are as good as the best performing one by extending the sequence.
%We have also conducted initial experiments with the t$|$ket$\rangle$ and Cirq compiler and come to the same conclusions.
%Even the corrected suboptimal sequence with three passes is at best equally good as the best performing two-pass sequence.
Experiments on multi-pass sequences show that more than two optimization passes may have an impact on each other but that this always links to the previously found pairwise effects.
We observe that it is important to initially choose the best order of optimization passes to get the best possible optimization for the given circuit. 
The experiments reveal some factors which are important to choose the best, or at least a good, order; among those, the resulting optimization sequence depends the most on the native gate set.

\end{abstract}


\copyrightnotice{}

\section{Introduction}\label{sec:intro}

Quantum computers are promising to solve certain problems faster than classical computers.
For example, Shor's~\cite{shor97} and Grover's algorithm~\cite{grover96} have a proven advantage to their classical counterparts.
To run quantum algorithms on quantum hardware, a quantum program must first be compiled to ensure it meets all constraints and is executable on the chosen device. 
Quantum compilation, including the whole procedure of processing the initial quantum circuit to the circuit that is run on the quantum hardware, usually induces significant overhead that leads to larger and deeper quantum circuits, negatively affecting the results due to noise. 
This becomes even more critical in the context of high-performance computing and quantum computing (HPCQC) integration, where the scalability and fidelity of quantum subroutines directly impact the overall system performance~\cite{Seitz_HPCQC, HPCQCchallenges}.
Reviews on quantum programming tools have highlighted the importance of compilers in HPCQC integration, and the need for robust optimization strategies~\cite{HPCQCReview}.
%Alongside software-level approaches like compiler optimization, dedicated hardware accelerators, such as FPGA-based systems designed to speed up specific tasks like neutral atom rearrangement~\cite{FPGA}, also contribute to enhancing overall quantum system performance.

To make quantum programs less error-prone, quantum circuit optimization, a part of quantum compilation where the quantum circuit is optimized with respect to a given metric, becomes relevant. 
%There are many optimization methods~\cite{amy2014,rahman2015,nam2018automated,Iten_2022,staudacher2023,PhysRevA.102.022406, peham2023clifford} that can be used.
While it is clear that the set of applied optimization passes affects the results, it is not yet clear whether the order of passes affects the results and, if it does, what the best order is. 
Current quantum compilers~\cite{qiskit, bergholm2022pennylaneautomaticdifferentiationhybrid, Cirq_Developers_2024, sivarajah_tket_2020, bqskit, cudaq, kim2023cudaq} typically use a fixed order for optimization passes but it is unclear whether the chosen order is the one that performs best or if changing the order makes a difference.
The question arises whether developers can arbitrarily choose the order of optimizations or if there are patterns that give a recommendation which order of optimization passes results in the best possible, or at least a good, quantum circuit w.r.t. a given metric.

In this paper, we analyze whether the order of optimization passes makes a difference for the resulting optimized circuit and which factors influence this. 
We focus on optimizations in the native gate set before mapping and evaluate pairwise combinations of optimization methods.
The results show that the order has an effect on the resulting circuit and that there are some pairwise dependencies.
Motivated by that, we examine whether repeating passes to correct non-optimal optimization sequences improves the outcome, which indeed does in some cases.
We investigate how pairwise interactions influence the optimization outcome using pass sequences consisting of more than two optimization passes.
Our results indicate that there are multi-pass interactions, but the pairwise interactions always occur when there is a difference in the resulting circuit between orders.

In the following, the basic concepts of quantum compilers, gate sets, and quantum circuit optimization are explained in Section~\ref{sec:background}. 
Section~\ref{sec:relwork} covers a rough overview to related research on the order of optimization passes.
The experiment setup is presented in Section~\ref{sec:implementation}, and the results are presented in Section~\ref{sec:results} and discussed in Section~\ref{sec:dicussion}.
The work is concluded and future perspectives are proposed in Section~\ref{sec:conclusion_fw}.

\section{Background}\label{sec:background}

Quantum compilers bridge the gap between high-level quantum algorithms and executable instructions for quantum hardware, typically through multiple compilation stages that optimize quantum circuits for efficient execution~\cite{Chong2017,UQP}, a critical aspect as efforts increase to integrate quantum accelerators into HPC systems via unified platforms~\cite{BridgeTheGap}.
Quantum computations are described in terms of quantum gates, fundamental operations on qubits, organized within circuits.
These gates often come from a universal, hardware-independent set, facilitating portability and abstraction. 
Quantum devices typically %support only a limited, hardware-specific gate set, also known as 
only support their native gate set, requiring transforming universal gates into these hardware-specific gates.
Quantum compilers implement both target-agnostic optimizations and target-dependent optimizations. 

\subsection{Quantum Circuit Optimization in Quantum Compilation}\label{subsec:quantum-comp}

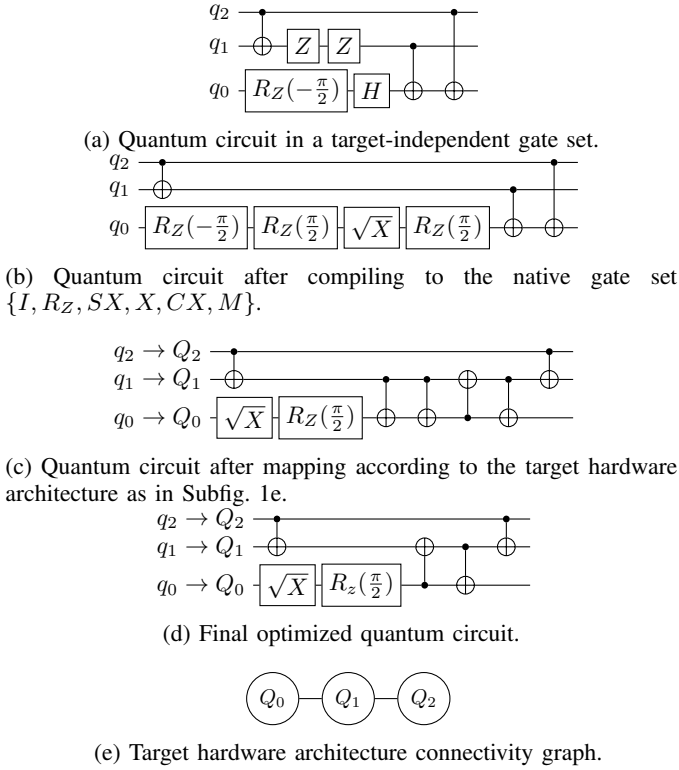
\begin{figure}
    \centering
    \begin{subfigure}{0.49\textwidth}
        \centering
        \begin{tikzpicture}[scale=0.9]
            \begin{yquant}
                qubit {$q_2$} q2[1];
                qubit {$q_1$} q1[1];
                qubit {$q_0$} q0[1];
                cnot q1[0] | q2[0];
                z q1[0];
                z q1[0];
                box {$R_Z(-\frac{\pi}{2})$} q0[0];
                h q0[0];
                cnot q0[0] | q1[0];
                cnot q0[0] | q2[0];
            \end{yquant}
        \end{tikzpicture}
        \caption{Quantum circuit in a target-independent gate set.}
        \label{subfig:ex-compilation-targetindep}
    \end{subfigure}
    \begin{subfigure}{0.49\textwidth}
        \centering
        \begin{tikzpicture}[scale=0.9]
            \begin{yquant}
                qubit {$q_2$} q2[1];
                qubit {$q_1$} q1[1];
                qubit {$q_0$} q0[1];
                cnot q1[0] | q2[0];
                box {$R_Z(-\frac{\pi}{2})$} q0[0];
                box {$R_Z(\frac{\pi}{2})$} q0[0];
                box {$\sqrt{X}$} q0[0];
                box {$R_Z(\frac{\pi}{2})$} q0[0];
                cnot q0[0] | q1[0];
                cnot q0[0] | q2[0];
            \end{yquant}
        \end{tikzpicture}
        \caption{Quantum circuit after compiling to the native gate set $\{I, R_Z, SX, X, CX, M\}$.}
        \label{subfig:ex-compilation-targetdep}
    \end{subfigure}
    \\[0.3cm]
    \begin{subfigure}{0.49\textwidth}
        \centering
        \begin{tikzpicture}[scale=0.9]
            \begin{yquant}
                qubit {$q_2\rightarrow Q_2$} q2[1];
                qubit {$q_1\rightarrow Q_1$} q1[1];
                qubit {$q_0\rightarrow Q_0$} q0[1];
                cnot q1[0] | q2[0];
                box {$\sqrt{X}$} q0[0];
                box {$R_Z(\frac{\pi}{2})$} q0[0];
                cnot q0[0] | q1[0];
                cnot q0[0] | q1[0];
                cnot q1[0] | q0[0];
                cnot q0[0] | q1[0];
                cnot q1[0] | q2[0];
            \end{yquant}
        \end{tikzpicture}
        \caption{Quantum circuit after mapping according to the target hardware architecture as in Subfig.~\ref{subfig:ex-compilation-hw-arch}.}
        \label{subfig:ex-compilation-target-architecture}
    \end{subfigure}
    \begin{subfigure}{0.49\textwidth}
        \centering
        \begin{tikzpicture}[scale=0.9]
            \begin{yquant}
                qubit {$q_2\rightarrow Q_2$} q2[1];
                qubit {$q_1\rightarrow Q_1$} q1[1];
                qubit {$q_0\rightarrow Q_0$} q0[1];
                cnot q1[0] | q2[0];
                box {$\sqrt{X}$} q0[0];
                box {$R_z(\frac{\pi}{2})$} q0[0];
                cnot q1[0] | q0[0];
                cnot q0[0] | q1[0];
                cnot q1[0] | q2[0];
            \end{yquant}
        \end{tikzpicture}
        \caption{Final optimized quantum circuit.}
        \label{subfig:ex-compilation-finalopt}
    \end{subfigure}
    \\[0.3cm]
    \begin{subfigure}{0.5\textwidth}
        \centering
        \begin{tikzpicture}[scale=0.5]
            \node[shape=circle,draw=black, scale=0.8] (A) at (0,0) {$Q_0$};
            \node[shape=circle,draw=black, scale=0.8] (B) at (2,0) {$Q_1$};
            \node[shape=circle,draw=black, scale=0.8] (C) at (4,0) {$Q_2$};
            \draw [-]
                (A) edge (B)
                (B) edge (C);
        \end{tikzpicture}
        \caption{Target hardware architecture connectivity graph.}
        \label{subfig:ex-compilation-hw-arch}
    \end{subfigure}
    \caption{Phases a quantum circuit goes through during compilation. Subfigs.~\ref{subfig:ex-compilation-targetindep} and~\ref{subfig:ex-compilation-targetdep} show the quantum circuits in the initial stage and after translation to the native gate set respectively. The connectivity graph used for mapping is depicted in Subfig.~\ref{subfig:ex-compilation-hw-arch}. 
    The circuit after mapping is shown in Subfig.~\ref{subfig:ex-compilation-target-architecture} and Subfig.~\ref{subfig:ex-compilation-finalopt} displays the final optimized circuit.}
    \label{fig:ex-compilation}
\end{figure}

Fig.~\ref{fig:ex-compilation} illustrates the phases a quantum circuit goes through during quantum compilation with an example circuit.
Subfig.~\ref{subfig:ex-compilation-targetindep} depicts the initial target-independent circuit.
The quantum circuit can be optimized directly in this stage using hardware-independent optimization methods.
Then the circuit is translated to the native gate set, in this example: $\{I, R_Z, SX, X, CX, M\}$ ($M$: measurement).
The resulting circuit is displayed in Subfig.~\ref{subfig:ex-compilation-targetdep}.
In this stage, the circuit can again be optimized using target-specific optimization methods.
Afterwards, the circuit is adapted to the connectivity of the target hardware with mapping, consisting of layouting (assigning the qubits in the quantum circuit to the qubits on the quantum device) and routing (placing and moving qubits to enable multi-qubit operations).
The connectivity graph used in the example is shown in Subfig.~\ref{subfig:ex-compilation-hw-arch}, and the circuit after mapping is shown in Subfig.~\ref{subfig:ex-compilation-target-architecture}.
The quantum circuit can then be optimized again using target-specific optimizations, resulting in the final optimized circuit in Subfig.~\ref{subfig:ex-compilation-finalopt}.

\subsection{Quantum Circuit Optimization}\label{subsec:circ-opt}

Quantum circuit optimization comprises all kinds of optimizations~\cite{amy2014,rahman2015,nam2018automated,Iten_2022,staudacher2023,PhysRevA.102.022406, peham2023clifford} that are conducted on a quantum circuit to make it more efficient according to the given metric(s).
Quantum circuit optimizations can generally be classified into different categories.
One possible classification is: pattern matching (pattern-based circuit improvements), gate cancellation (removing redundant gate operations), Hadamard gate reduction (reducing the Hadamard gate count), Merging $R_Z$ gates, and depth reduction (minimizing the longest sequence of sequential gates)~\cite{quantum7010002}.
Another classification is: decomposition, reduction, commutation, and structural optimizations~\cite{swierkowska24}.
Some optimization methods can also be a combination of multiple categories.

Common metrics to evaluate quantum circuit optimization techniques include, for instance, single-qubit and two-qubit gate count, circuit depth, and fidelity estimation, which directly influence the reliability, execution speed, and resource requirements of quantum and classical computations.

\section{Related Work}\label{sec:relwork}

Current literature on quantum circuit optimization mostly concerns itself with optimization methods themselves, where the goal is to find an approach that minimizes the circuit regarding a specific metric.
To the best of our knowledge, there has not been much research on how optimization methods influence each other during compilation.

\subsection{Optimization in Classical Compilers}\label{subsec:classical-compilers}

In classical computing, compilers also use various optimizations to improve the performance of a program.
Finding the optimal order of optimization passes in this context is known as the \emph{phase ordering problem}~\cite{huang2020autophasejugglinghlsphase,touati2006}.
Current classical compilers %, such as LLVM~\cite{LLVM} and GCC~\cite{gcc},
employ optimization techniques analogous to quantum compilers, using predefined orders of optimization passes to enhance program efficiency and execution performance.
For instance, in LLVM~\cite{LLVM} the order of passes at each level is chosen based on extensive empirical evaluation to balance compile time overhead against run time performance improvements; GCC’s~\cite{gcc} pass ordering is tuned by compiler developers through extensive benchmarking and continuous testing.
The fixed order of passes is not optimal for every program~\cite{Almakki_Ali_Izzeldin_Huang_Cummins}.
It has been proven that in the classical domain, the phase ordering problem is generally undecidable, but for simplified cases, the problem becomes decidable~\cite{touati2006}.

These significant differences in classical compilers underscore the necessity of systematically exploring the phase ordering problem in quantum compilation.

\subsection{Circuit Optimization Order in Quantum Compilers}\label{subsec:sota-compilers}

Quantum platform providers mostly also provide a compiler with which a submitted quantum circuit is processed.
These compilers take the task of quantum circuit optimization, where multiple optimization passes are sequentially applied on the circuit to get a better one regarding a given metric.

For instance, \emph{Qiskit's}~\cite{qiskit} compiler optimizes the circuit regarding the circuit depth, single-qubit gate count, and two-qubit gate count after the translation to the native gate set.
The order of the passes is predefined and the extent of optimization depends on the optimization level, going from level 0 (no optimization) to level 3 (highest level of optimization).
In \emph{PennyLane}~\cite{bergholm2022pennylaneautomaticdifferentiationhybrid}, the available optimization passes mainly focus on reducing the circuit depth and also work on the native gate set if specified.
A default pass application order is used thereby. 
%First, all commuting single-qubit gates are moved to the right. 
%Then all neighboring gates that are inverse to each other are canceled, and all neighboring rotation gates of the same type are merged together.
In current quantum compilers~\cite{qiskit, bergholm2022pennylaneautomaticdifferentiationhybrid, Cirq_Developers_2024, sivarajah_tket_2020, bqskit, cudaq, kim2023cudaq}, the order of optimization passes is usually fixed by default.
% The optimizations that are run by default in \emph{Cirq}~\cite{Cirq_Developers_2024} are generally separated in three steps: pre-processing, translation to the native gate set, and post-processing.
% The steps are identical to the ones described in Subsection~\ref{subsec:quantum-comp}, Fig.~\ref{fig:ex-compilation}, up to the step of mapping (excluded).
Apart from the aforementioned compilers, various other ones have been proposed (see, e.g.,~\cite{sivarajah_tket_2020, bqskit, cudaq, kim2023cudaq, Cirq_Developers_2024}).

\subsection{Optimization Pass Selection}

Closely related to the phase ordering problem is the selection of the best optimizations for the specific quantum circuit.
There are approaches that address this problem using heuristics.
One example is the MQT Predictor~\cite{quetschlich2024mqtpredictor} that uses reinforcement learning to find the best set of passes for a device-specific quantum compiler.
It combines multiple passes from different compiler tools to find the best passes for the given use case.
%The authors used \emph{expected fidelity} or \emph{circuit depth} as metrics and t
The results show that the MQT Predictor creates circuits that are in most cases better than any tested combination.
Other approaches~\cite{swierkowska24,dangwal2025cliffordassistedoptimalpass} include solutions that use genetic algorithms such as the Non-dominated Sorting Genetic Algorithm II (NSGA-II)~\cite{deb2000} or dummy circuits that are similar to the quantum program with a divide-and-conquer approach (the tool realizing this is called E-COMPASS~\cite{dangwal2025cliffordassistedoptimalpass}).

Although these approaches %such as MQT Predictor and E-COMPASS 
advance pass selection, they do not systematically address the phase ordering problem in the quantum realm. 
Understanding pass dependencies and interactions remains essential because even optimal pass selection can underperform without the correct order, motivating our exploration of the phase ordering problem.

\section{Experimental Setup}\label{sec:implementation}
%\RTWO{Runtime and computational cost are largely absent from the discussion.  Since the number of combinations grows rapidly, reporting the experimental cost, compile times, and scalability limitations would make the study more complete.\\ -- I can mention that it took days to finish an experiment and that the run time scales very badly when the number of qubits rises}

The goal of this work is to get insights into the impact of the order of optimization passes on the resulting optimized quantum circuit.
The idea is to evaluate various different pass combinations and compare the resulting circuits of the respective permutations.
For this, we look into optimizations targeting the native gate set before mapping.
We exclude mapping since it is highly dependent on the topology of the quantum hardware the circuit should be run on and we follow a more general approach to better examine interactions between optimization passes.

We define and systematically evaluate optimization pass sequences in stages: 
first, we analyze pairwise relations to detect the influence they have on each other, then examine if it is possible to correct less performing sequences, and finally expand the analysis to multi-pass sequences. 
This structured approach ensures comprehensive insights into how two optimization passes influence each other.

We base our experiments on Qiskit v2.3.0 and use its optimization passes.
They were conducted multiple times and show similar results in each repetition, suggesting that the results are deterministic.
In the following subsections, the choice of optimization passes is explained in Section~\ref{subsec:opt_passes}.
The metrics chosen for evaluation are described in Section~\ref{subsec:metrics} and the benchmarks that are used are listed in Section~\ref{subsec:benchmarks}.

\subsection{Optimization Passes}\label{subsec:opt_passes}

The experiments use Qiskit's predefined optimization passes, which are separated into two types: transformation passes and analysis passes.
We only consider transformation passes for the experiments, since analysis passes do not actively change the structure of the quantum circuit, but analyze it for specific criteria used by transformation passes.
If a transformation pass requires information that is provided by an analysis pass, the compiler automatically adds them to the optimization sequence.
Collect-and-collapse passes, a subtype of the transformation passes, are excluded since they only replace certain blocks of gates with a more general gate that is usually neither supported by hardware-independent gate sets nor by native gate sets; passes that necessitate further information on the specific quantum hardware are also excluded. % HIER Info hardwar-independent rauslassen wäre okay
Hence, out of 32 predefined optimization passes, the following 16 are selected for the experiments:
\begin{itemize}
    \item \emph{Op\-ti\-mize\-1q\-Gates}: Merge multiple single-qubit gates into one single gate
    \item \emph{Op\-ti\-mize\-1q\-Gates\-De\-com\-po\-si\-tion}: Merge multiple single-qubit gates into a simplified gate sequence
    \item \emph{In\-verse\-Can\-cel\-la\-tion}: Cancel inverse gate pairs
    \item \emph{Com\-mu\-ta\-tive\-Can\-cel\-la\-tion}: Cancel self-adjoint gates ($H, X, Y, Z, CX, CY, CZ$) using commutation relations
    \item \emph{Com\-mu\-ta\-tive\-In\-verse\-Can\-cel\-la\-tion}: Cancel inverse gate pairs using commutation relations
    \item \emph{Op\-ti\-mize\-1q\-Gates\-Simple\-Com\-mu\-ta\-tion}: Commute single-qubit gates through two-qubit gates when possible, then do the same as \textit{Op\-ti\-mize\-1q\-Gates\-De\-com\-po\-si\-tion}
    \item \emph{Con\-tract\-I\-dle\-Wires\-In\-Con\-trol\-Flow}: Remove unused qubits from (classical) control-flow operations 
    \item \emph{Re\-move\-Di\-ag\-o\-nal\-Gates\-Be\-fore\-Mea\-sure}: Remove diagonal gates before a measurement
    \item \emph{Re\-move\-Re\-set\-In\-Zero\-State}: Remove the reset gate when the qubit is in the ``zero state''
    \item \emph{Tem\-plate\-Op\-ti\-miza\-tion}: Replace certain patterns with a simplified equivalent
    \item \emph{Re\-set\-Af\-ter\-Mea\-sure\-Sim\-pli\-fi\-ca\-tion}: Replace the reset gate after measurements with a simplified gate
    \item \emph{Op\-ti\-mize\-Clif\-fords}: Merge multiple Clifford gates
    \item \emph{E\-lide\-Per\-mu\-ta\-tions}: Remove SWAP and permutation gates and permute the virtual qubit positions
    \item \emph{Op\-ti\-mize\-An\-no\-tat\-ed}: Optimize the circuit according to given gate annotations 
    \item \emph{Split\-2Q\-U\-ni\-ta\-ries}: Split two-qubit unitaries to two single-qubit gates
    \item \emph{Re\-move\-I\-den\-ti\-ty\-E\-qui\-va\-lent}: Remove gates close to the identity operation (i.e., gates with a negligible effect)
\end{itemize}

We configure these with minimal parameters to not restrict the optimizations, and retranslate the circuit to the predefined native gate set after all optimizations have been applied.
The chosen optimization passes do not always improve the circuit by themselves regarding the considered metrics but it is possible that they do once more optimization passes are added to the optimization sequence.
We are mainly interested in whether the order of optimization passes makes a difference, so we take all into account independent of the performance.

To determine whether the optimization passes have an impact on each other, different permutations of the same set of optimization passes are tested and compared.
For this, the 16 chosen optimization passes are paired with each other, respectively.
%; the total number of pass combinations $c$ for $k$-pass sequences with $n$ passes in total is calculated with the binomial coefficient $c=\binom{n}{k}$.
These combinations are permuted and, since there are only two passes, there are two possible permutations for each combination.
%Hence, according to the urn model~\cite{urnmodel}, the total number of optimization sequences $s$ to test is $s=k!\cdot c=k!\cdot\binom{n}{k}$.
We first focus on pairwise combinations, so we use $c=\binom{16}{2}=120$ combinations and test a total number of $s=2!\cdot\binom{16}{2}=240$ sequences for each benchmark (of different sizes) and native gate set combination.

\subsection{Metrics}\label{subsec:metrics}
%\RONE{Depth and gate count are useful metrics, but the study would be more complete with at least some discussion of compilation runtime (e.g., gate runtimes can skew the phase ordering if different pass orders has biased gate-reduction performance), fidelity-oriented metrics, or downstream hardware execution implications. \\ -- Mention that in future work, or mention that this is out of scope because this takes specific hardware into consideration}

To evaluate whether different orders of optimization passes influence each other, we compare them using the metrics circuit depth (cd) and gate count (gc).
These are the most straightforward metrics to compare how well an optimization method works.
Initially, we split the gate count into single-qubit gates and two-qubit gates. 
First experiments show that the used optimization passes do not optimize the two-qubit gate count in almost all instances and therefore do not show any differences in the resulting optimized circuit.
Hence, we combine single-qubit and two-qubit gate count to the overall gate count.

To compare orders of optimization passes, we compare how well they optimize the initial circuit.
It is calculated as follows:
$$R_{m}(p)=\frac{C_{m}(x_p)}{C_{m}(x_{init})}*100\%$$
where $C_{m}(x)$ is the cost function that denotes the value of the metric $m\in$ \{cd, gc\} for the quantum circuit $x$. The parameter $x_p$ is the circuit after applying the order of optimization passes $p$ and the parameter $x_{init}$ denotes the initial non-optimized quantum circuit.
This means that $R_{m}(p)\geq 100\%$ shows that the optimization passes using this order do not improve the circuit, while with $R_{m}(p)<100\%$, the optimization passes in this order improve the circuit.
The lower $R_{m}(p)$, the better the optimization.
We then compare the respective values of two different orders of optimization passes $p_0,p_1$ by comparing $R_{m}(p_0)$ with $R_{m}(p_1)$.

\subsection{Benchmark Circuits}\label{subsec:benchmarks}

As a benchmark, we use quantum circuits provided by MQT Bench~\cite{quetschlich2023mqtbench} and choose the circuits amplitude estimation (ae), the Deutsch-Josza algorithm (dj), graph state (graphstate), quantum Fourier transform (qft), a random (but fixed) circuit (random), and VQE two local ansatz (vqe\_two\_local).
These benchmarks represent different classes of circuits: 
Amplitude estimation represents a circuit solving a given problem including the quantum Fourier transform, the graph state is a graph representation using quantum circuits, the Deutsch-Josza algorithm is an oracle algorithm, the quantum Fourier transform circuit shows the pure QFT, the VQE two local ansatz uses a parameterized circuit with random parameters (representing variational quantum algorithms), and the random circuit is a circuit that might reveal results that are not obvious using the other, more structured, benchmark circuits.

Initial investigations show that most optimization passes work better with the native gate set than the hardware-independent gate set, which is the reason for omitting optimization in the hardware-independent gate set and only optimizing circuits represented in the native gate set.
The native gate sets considered in this work are taken from those provided by MQT Bench~\cite{quetschlich2023mqtbench} and are the following:
\begin{center}
\small
    \begin{tabular}{|c|c|}
        \hline
        Name & Native Gate Set \\\hline\hline
        IBM Falcon & $I$, $X$, $SX$, $R_Z$, $CX$, $M$ \\\hline
        IonQ Forte & $R_Z$, $GPi$, $GPi2$, $M$ \\\hline
        IQM & $R$, $CZ$ \\\hline
        Quantinuum & $R_X$, $R_Y$, $R_Z$, $R_{ZZ}$\\\hline
        Rigetti & $R_X(\pi)$, $R_X(\frac{\pi}{2})$, $R_X(\frac{\pi}{2})^\dagger$, $RZ$, $ISWAP$, $M$ \\\hline
    \end{tabular}
\end{center}
The gate $M$ denotes the measurement.
We conduct the experiments with a benchmark circuit size ranging from five to twelve qubits. 
For time and memory reasons, the experiments that include the \emph{Tem\-plate\-Op\-ti\-miza\-tion} pass only go up to ten qubits.
With an increasing benchmark size, the run time of the experiments increases significantly and goes from taking minutes to taking days for one run of all sequences for a benchmark - benchmark size - native gate set combination.
The chosen circuit sizes are rather small, but are enough to demonstrate the main findings.
%While larger circuits may exhibit additional effects, we expect the observed differences to be similar to our observed ones, as the experiments show that increasing the number of qubits does not influence whether there is a difference between permutations of a pass sequence. 
We expect that with larger circuits the results become even more pronounced, as increased circuit size provides greater opportunity for each optimization pass to take effect, but we do not expect the results to change with a higher number of qubits.

\section{Experimental Results}\label{sec:results}
%\RONE{The current evaluation is centered on Qiskit, so the paper would be stronger if it clarified which findings are likely compiler-agnostic and which may depend on specific implementation details of that SDK (.e.g., MLIR). \\ -- Mention that we have done initial experiments on tket and Cirq and saw that the findings are similar}
%\RTWO{Since the conclusions are derived entirely from experiments, more attention should be paid to reproducibility and variability.  Qiskit's optimization level 3 may contain non-deterministic components, and the paper does not appear to repeat experiments multiple times.  Given the large search space, this omission is understandable, but the possibility of non-determinism and its implications should be explicitly acknowledged.  Ideally, the authors should repeat a representative subset of experiments several times and report whether the observed differences are statistically stable. \\ -- Mention that the experiments were repeated multiple times but the dependencies strongly depend on the implementation}

In Subsection~\ref{subsec:pairwise_deps}, an analysis on pairwise combinations and the influence of the order of optimization passes is conducted.
The results show that there is a difference when permuting the order of passes in some cases; following that, the question arises whether a suboptimal order can be ``corrected'', which is then investigated in Subsection~\ref{subsec:correct_passorder}.
To examine how pairwise interactions and order permutations affect longer pass sequences, we conduct experiments with more than two optimization passes, detailed in Subsection~\ref{subsec:multipass_deps}.

We use the notation that the better one of the two possible optimization pass sequences, i.e., the better permutation, is called $p_b$, and the worse one $p_w$.

\subsection{Pairwise Combinations}\label{subsec:pairwise_deps}

Experiments on pairwise combinations of optimization passes show that 12 out of the 120 combinations have a difference between the two possible sequence orders.
Table~\ref{tab:comb_deps} shows the sequences and differences for these combination pairs. 
The first column shows the IDs we use for the following sections to refer to the respective pass pairs, the second column shows $p_b$ of each pair, i.e., the better optimization order, and the next columns show the minimum, maximum, mean, and median differences between the two permutations for the circuit depth and gate count.
The differences are computed with $R_{m}(p_w)-R_{m}(p_b)$ meaning that the greater the value in the table, the better $p_b$ is compared to $p_w$.
Negative values in the table indicate that there are instances where $p_w$ results in a better circuit than $p_b$.

\begin{table*}[]
    \centering
    \small
    \caption{Differences between permutations of those optimization pass pair combinations where the order of passes has an impact on the resulting optimized circuit. The differences are computed with $R_{m}(p_w)-R_{m}(p_b)$ and the minimum, maximum, mean, and median values are given for the circuit depth and the gate count.}
    \addtolength{\tabcolsep}{-0.3em}
    \begin{tabular}{|c|c|rrr|rrr|}
        \hline
         \multirow{2}{*}{ID} & \multirow{2}{*}{Optimization Pass Pair} & \multicolumn{3}{r|}{Difference Circuit Depth} & \multicolumn{3}{r|}{Difference Gate Count} \\
         & & Min. & Max. & Mean/Med. & Min. & Max. & Mean/Med. \\ \hline\hline
         1 & [Optimize1qGatesSimpleCommutation, Optimize1qGates] & 0 & 134.99 & 63.93/57.38 & 0 & 125.71 & 62.07/55.07 \\\hline %4
         2 & [CommutativeCancellation, Optimize1qGatesDecomposition] & -1.00 & 10.53 & 3.45/-0.04 & -10.04 & 10.31 & 0.63/0.06 \\\hline %15
         3 & [Optimize1qGatesDecomposition, CommutativeInverseCancellation] & -1.00 & 0 & -1.00/-1.00 & -2.18 & 1.25 & 0.22/0.13 \\\hline %16
         4 & [Optimize1qGatesDecomposition, Optimize1qGatesSimpleCommutation] & 0 & 134.99 & 63.7/57.38 & 0 & 125.71 & 62.19/54.55 \\\hline %17
         5 & [TemplateOptimization, Optimize1qGatesDecomposition] & -0.68 & 134.92 & 20.28/7.69 & 0 & 51.76 & 10.4/7.14 \\\hline %20
         6 & [CommutativeInverseCancellation, CommutativeCancellation] & 0.19 & 2.63 & 0.72/0.13 & -0.13 & 0.74 & 0.16/0.08 \\\hline %39
         7 & [CommutativeCancellation, Optimize1qGatesSimpleCommutation] & -3.92 & 121.55 & 84.56/113.68 & -0.05 & 108.03 & 55.55/97.47 \\\hline %40
         8 & [TemplateOptimization, CommutativeCancellation] & 0 & 27.59 & 6.8/3.02 & 0 & 28.15 & 5.38/2.59 \\\hline %43
         9 & [CommutativeInverseCancellation, Optimize1qGatesSimpleCommutation] & 0 & 134.71 & 100.64/95.35 & -0.08 & 107.43 & 60.28/81.15 \\\hline %50
         10 & [TemplateOptimization, Optimize1qGatesSimpleCommutation] & -6.86 & 267.76 & 36.28/22.5 & -18.42 & 77.37 & 22.39/13.92 \\\hline %62
         11 & [RemoveIdentityEquivalent, Optimize1qGatesSimpleCommutation] & 0 & 113.88 & 78.72/71.35 & 0 & 92.9 & 65.05/68.87 \\\hline %68
         12 & [TemplateOptimization, RemoveDiagonalGatesBeforeMeasure] & 0 & 0.19 & 0.12/0.11 & 0 & 0.06 & 0.03/0.02 \\\hline %70
    \end{tabular}
    \label{tab:comb_deps}
\end{table*}
%%%%%%%%%%%%%%%%%%
%\RTWO{The current presentation of Table I is difficult to read. All numerical values should be right-aligned and reported with a fixed precision (e.g., two decimal places).  In addition, the best values and preferred pass combinations should be highlighted in boldface to improve readability and facilitate comparison. \\ -- Do that!}
%%%%%%%%%%%%%%%%%%

Out of the 16 optimization passes, the following set of eight passes exhibits pairwise relations: \{\emph{Op\-ti\-mize\-1q\-Gates}, \emph{Op\-ti\-mize\-1q\-Gates\-De\-com\-po\-si\-tion}, \emph{Com\-mu\-ta\-tive\-Can\-cel\-la\-tion}, \emph{Com\-mu\-ta\-tive\-In\-verse\-Can\-cel\-la\-tion}, \emph{Optimize\-1q\-Gates\-Simple\-Com\-mutation}, \emph{Re\-move\-Di\-ag\-o\-nal\-Gates\-Be\-fore\-Mea\-sure}, \emph{Tem\-plate\-Op\-ti\-miza\-tion}, \emph{Re\-move\-I\-den\-ti\-ty\-E\-qui\-va\-lent}\}.

The table reveals that for most combinations the minimum difference is at least around $0$, meaning that the sequence $p_b$ in most cases performs at least equally well as $p_w$, if not better.
There are exceptions: In some cases the minimum value is below $0$, which means that there are single cases where $p_b$ is worse than $p_w$.
In general, the sequence $p_b$ is still better, since the mean and median values are much greater than $0$.

We find exceptions in two pairs, where the mean and median values are very close to $0$, meaning that there is not always an obvious better permutation.
One is ID 2: Overall, the minimum, maximum, mean and median tend to be positive and that means that there are more instances that are better using the determined $p_b$, so we deduce that $p_b$ is better.
The other one is ID 3: Regarding the depth, it seems that our assumed $p_w$ must be better than $p_b$ because the minimum, mean, and median are negative.
However, one must take into account that there is only one instance where the depth differs between the two permutations.
We hence focus on the gate count here, since there are more instances where a difference is observed, and choose the $p_b$ in our table as the better permutation for the same reason as for ID 2. 

\begin{figure}
    \centering
    \includegraphics[width=0.485\textwidth]{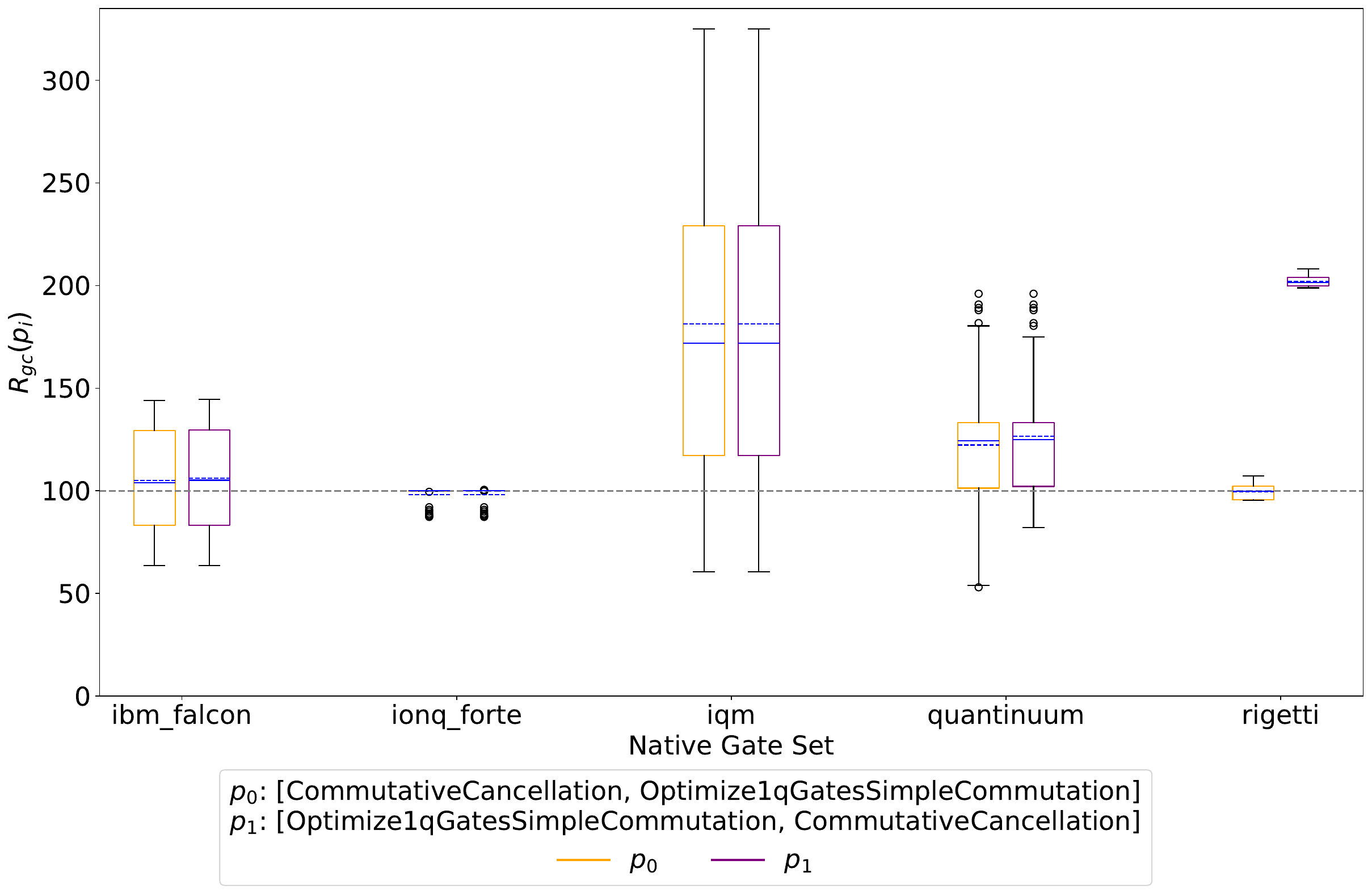}
    \caption{Optimization results when using the pass pair \{\emph{Com\-mu\-ta\-tive\-Can\-cel\-la\-tion}, \emph{Op\-ti\-mize\-1q\-Gates\-Simple\-Com\-mutation}\} compared to the non-optimized circuit. The used metric here is the gate count and the results are separated by benchmarks.}
    \label{fig:pairwise_ngs}
\end{figure}

We observe that the native gate set of the target device is decisive in some cases.
For instance, Fig.~\ref{fig:pairwise_ngs} holds the results for the pass pair \{\emph{Com\-mu\-ta\-tive\-Can\-cel\-la\-tion}, \emph{Op\-ti\-mize\-1q\-Gates\-Simple\-Com\-mutation}\} using the gate count as metric.
It displays the comparative value $R_{gc}(p_i)$ with $i\in\{0,1\}$ for $p_0$: [\emph{Com\-mu\-ta\-tive\-Can\-cel\-la\-tion}, \emph{Optimize\-1q\-Gates\-Simple\-Commutation}] and $p_1$: [\emph{Optimize\-1q\-Gates\-Simple\-Commutation}, \emph{Com\-mu\-ta\-tive\-Can\-cel\-la\-tion}] on the y-axis and the native gate set on the x-axis.
The left box plot (orange) for each native gate set shows $R_{gc}(p_0)$ and the right one (purple) shows $R_{gc}(p_1)$.
Dashed and solid lines (blue) in the box plots denote the mean and median values, respectively; the dashed baseline represents $R_{m}(p_i)=100\%$, i.e., the point where the resulting optimized circuit is equal to the non-optimized circuit.
Data below that line signify that the resulting optimized circuit got better and data above that line signify that the resulting optimized circuit got worse regarding the given metric $m$.

The figure shows that the differences between the permutations vary for different native gate sets: $R_{gc}(p_0)$ and $R_{gc}(p_1)$ differ the most for Rigetti's gate set. 
There are visible differences between the permutation results for the gate sets of IBM Falcon, IonQ Forte, and Quantinuum, but they are not as significant as with Rigetti's native gate set. 
With IQM's native gate set, changing the order of these two passes does not change the resulting circuit. 
Regarding the benchmarks, in some cases, for one benchmark, there is a bigger difference than for another, but it is not as apparent as when separating the resulting using the native gate set.

\begin{figure}
    \centering
    \includegraphics[width=0.485\textwidth]{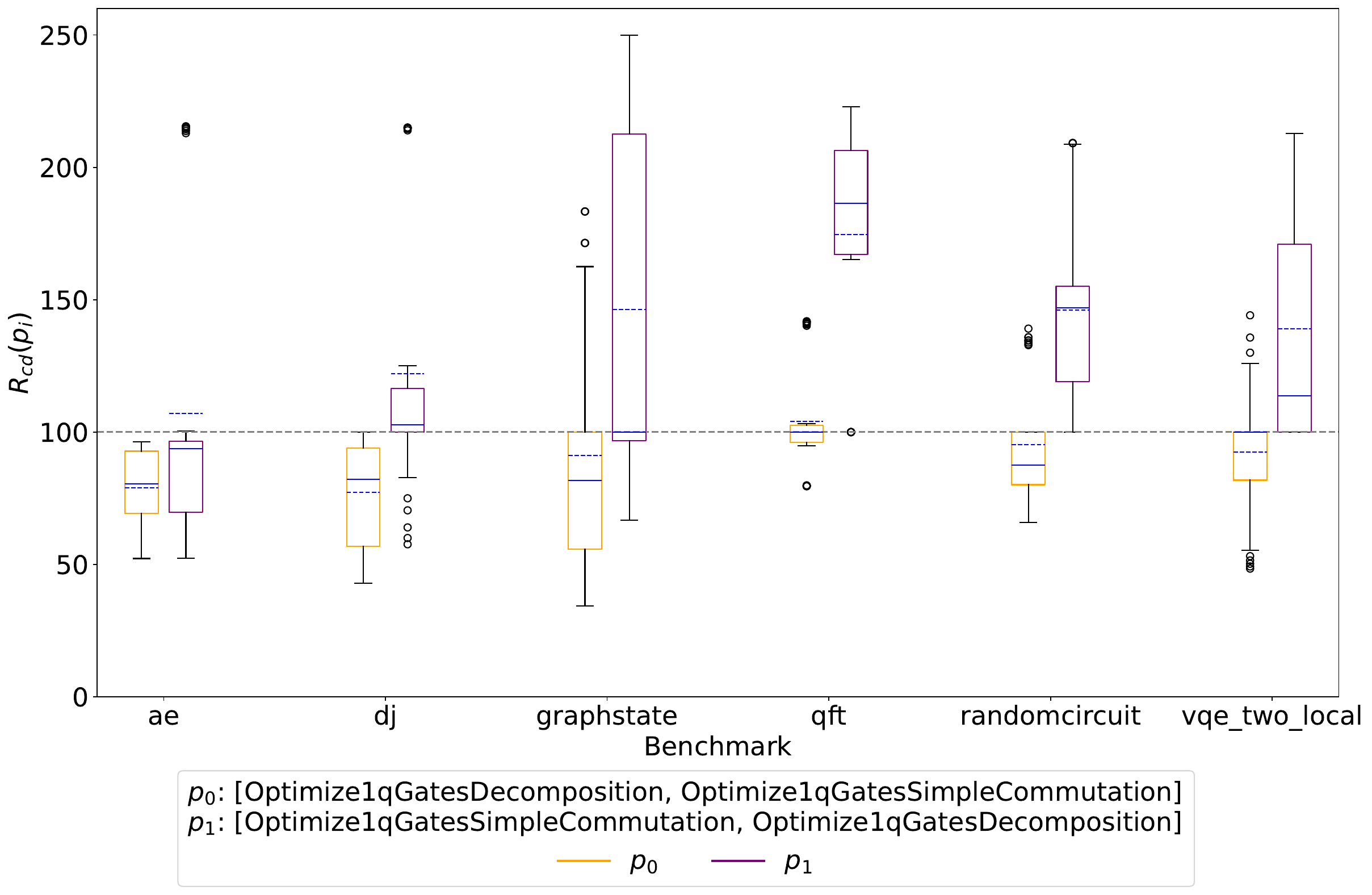}
    \caption{Optimization results when using the pass pair \{\emph{Op\-ti\-mize\-1q\-Gates\-De\-com\-po\-si\-tion}, \emph{Op\-ti\-mize\-1q\-Gates\-Simple\-Com\-mu\-ta\-tion}\} compared to the non-optimized circuit. Here, the gate count is the used metric and the results are separated by native gate set on the x-axis.}
    \label{fig:pairwise_bm}
\end{figure}

As can be seen in Fig.~\ref{fig:pairwise_ngs}, the optimizations do not always improve the circuit in comparison to the non-optimized one, but it can also be dependent on the order of optimizations how well the set of optimizations works.
%It is however possible that if there is a longer sequence of optimizations, they then output a better circuit than the non-optimized one.
Fig.~\ref{fig:pairwise_bm} demonstrates one example where the permutation makes a difference whether the circuit is improved.
This time, it depicts the results for the circuit depth and the benchmarks are on the x-axis.
The figure displays the results for the pass pair \{\emph{Op\-ti\-mize\-1q\-Gates\-De\-com\-po\-si\-tion}, \emph{Op\-ti\-mize\-1q\-Gates\-Simple\-Com\-mu\-ta\-tion}\}.
It makes clear that $p_0$ ([\emph{Op\-ti\-mize\-1q\-Gates\-De\-com\-po\-si\-tion}, \emph{Optimize\-1q\-Gates\-Simple\-Commutation}]) in most cases improves the circuit while applying the permutation $p_1$ ([\emph{Optimize\-1q\-Gates\-Simple\-Commutation}, \emph{Op\-ti\-mize\-1q\-Gates\-De\-com\-po\-si\-tion}]) results in a worse circuit w.r.t. the given metrics (we observe the same for the gate count).
The analysis of how well optimizations work in general in different scenarios is out of scope of this work.

\begin{tcolorbox}[mytakeaway]
The order of optimization passes makes a difference for the resulting optimized quantum circuit with respect to the given metrics circuit depth and gate count.
%It in some cases even makes a difference whether the circuit is optimized at all or gets worse than the non-optimized circuit.
There are clear tendencies which of the permutations is the best performing one, with two exceptions %: [Com\-mu\-ta\-tive\-Can\-cel\-la\-tion, Op\-ti\-mize\-1q\-Gates\-De\-com\-po\-si\-tion] and [Com\-mu\-ta\-tive\-In\-verse\-Can\-cel\-la\-tion, Op\-ti\-mize\-1q\-Gates\-De\-com\-po\-si\-tion], 
where it depends on the specific case. 
For these cases, there are light tendencies which order is better.
\end{tcolorbox}

\subsection{Correcting the Order of Passes}\label{subsec:correct_passorder}

After the observation that the order in which optimization passes are applied makes a difference for the resulting quantum circuit, the question arises whether a non-optimal order of passes can be ``corrected''.
We investigate this by checking whether the worse sequence $p_{w}$ is as good as the better one $p_{b}$ when reapplying the first optimization pass:
For instance, the optimization sequence [$o_0$, $o_1$] ($o_i$ $(i\in\{0,1\})$ are optimization passes) is best performing, but [$o_1$, $o_0$] was used for optimization.
Then $o_1$ is reapplied after $p_{w}$ such that the final sequence is $p_{w+}=[o_1, o_0, o_1]$, and therefore includes the best performing order of passes [$o_0$, $o_1$]. %, to find out if this results in the same optimized circuit as $p_{b}$, or an even better one.
This might then result in a circuit that is at least as good as the one resulting from $p_b$ since it is contained in the new sequence.
If this is the case, we say that the order of passes can be ``corrected''.
We also apply this process to the better performing sequence and use $p_{b+}=[o_0, o_1, o_0]$ to examine whether another application of the first optimization pass has an influence on the resulting circuit in this case. 

The experiment results reveal that five of the twelve optimization pass combinations of Table~\ref{tab:comb_deps} cannot be corrected (IDs 4, 6, 7, 9, and 11) and seven can be corrected (IDs 1, 2, 3, 5, 8, 10, and 12).
Note that not 100\% of the circuits is corrected in those seven cases, but in most instances. 
There are five cases where when appending the first optimization pass to the optimization sequence of the worse permutation, therefore using $p_{w+}$, the resulting optimized circuit is equally good as the circuit resulting from $p_b$.
Appending the first optimization pass to the better permutation, therefore using $p_{b+}$, does not change the resulting circuit.
% Fig.~\ref{fig:correct-order} shows an example plot for a sequence that can be corrected using the combination \{Op\-ti\-mize\-1q\-Gates, Optimize\-1q\-Gates\-Simple\-Commutation\}.
% The y-axis shows the values $R_m(p_i), i\in\{0,1\}$ between the permutations, and the benchmarks are on the x-axis.
% For each benchmark, there are four box plots for each sequence, from left to right: $p_0$, $p_1$, $p_{0+}$, and $p_{1+}$.
% The figure shows the results for the circuit depth; the results for the gate count are analogous to those.
% \begin{figure}
%     \centering
%     \includegraphics[width=0.485\textwidth]{figures/4_benchmarks_depth_.pdf}
%     \caption{TODO (Man könnte überlegen, hier die Benchmarks auch einfach zusammenzufassen) oder das generell wegzulassen}
%     \label{fig:correct-order}
% \end{figure}
% In this case, the permutation $p_1$ is generally better than $p_0$.
% The plots expose that, when adding the first optimization pass again to the optimization sequence of the worse permutation, therefore getting $p_{0+}$, the resulting optimized circuit is as good as the circuit resulting from $p_1$. 
% Appending the first optimization pass to the better permutation, therefore getting $p_{1+}$, in turn does not change the resulting circuit.

\begin{figure}
    \centering
    \includegraphics[width=0.485\textwidth]{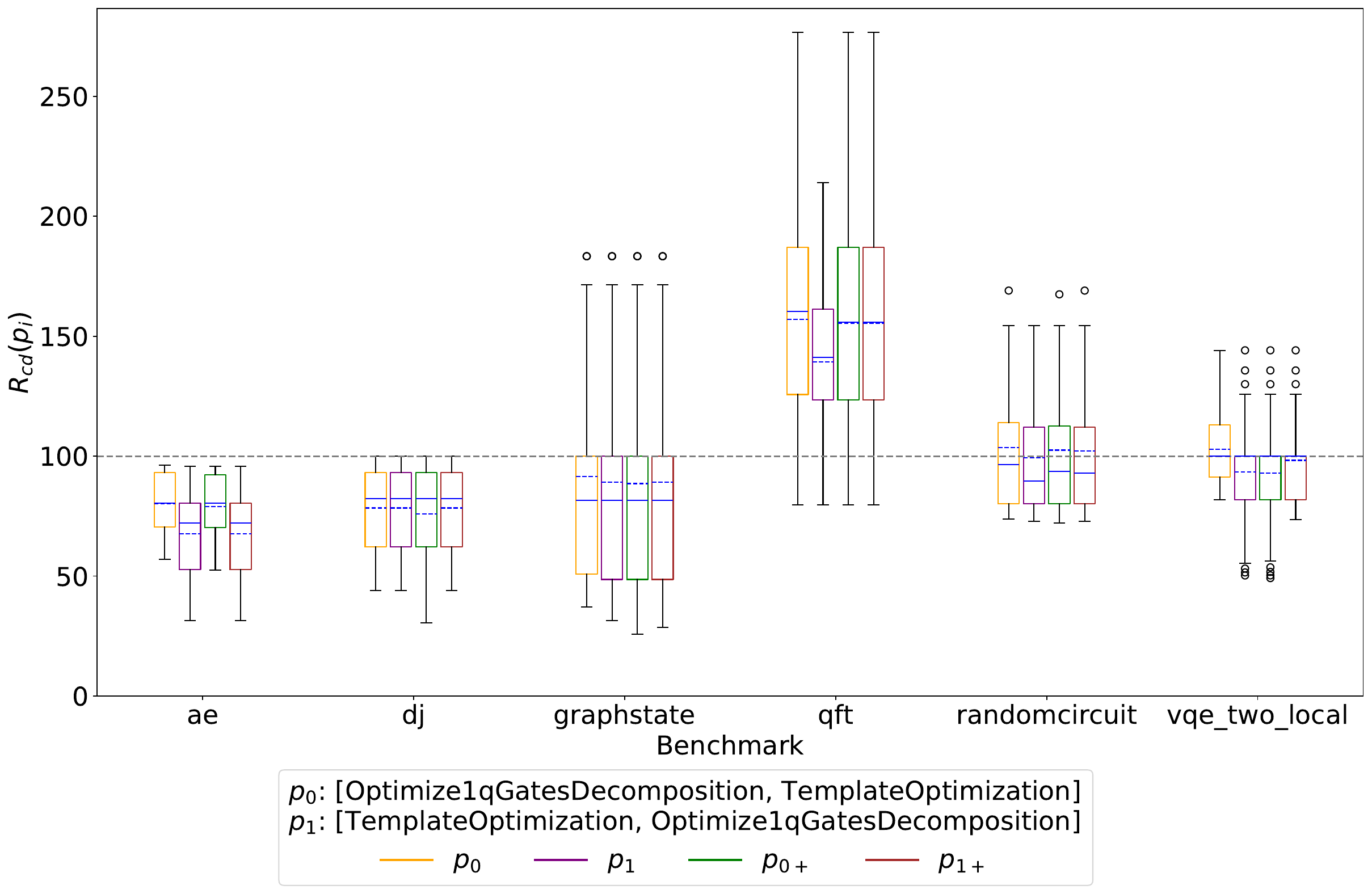}
    \caption{Results of conducting correction experiments for the pass pair \{\emph{Op\-ti\-mize\-1q\-Gates\-Simple\-Com\-mutation}, \emph{Tem\-plate\-Op\-ti\-miza\-tion}\}, separated by the benchmark on the x-axis. The circuit depth is used as metric.}
    \label{fig:improve-order-bm}
\end{figure}

\begin{figure}
    \centering
    \includegraphics[width=0.485\textwidth]{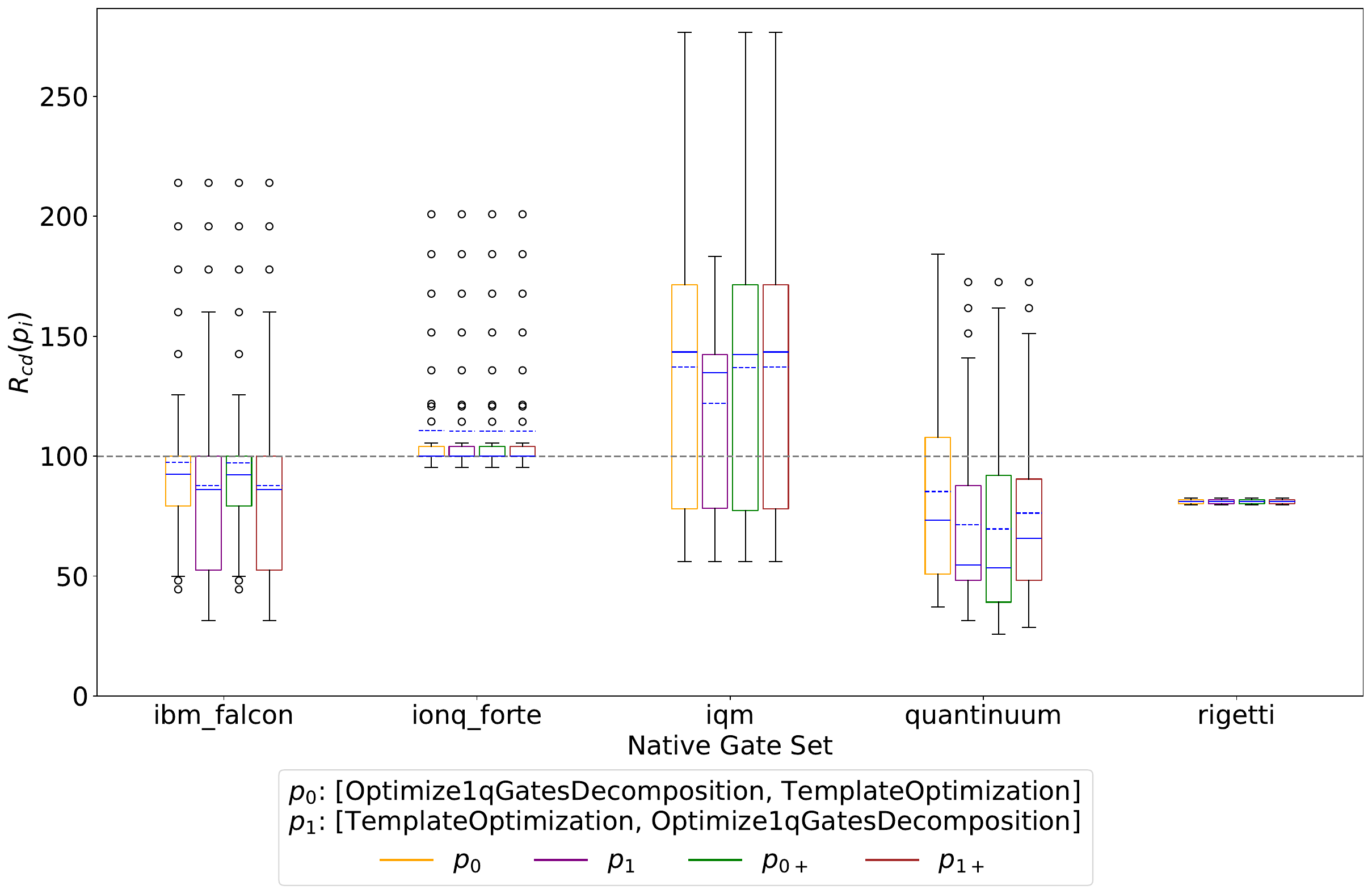}
    \caption{Results of conducting correction experiments for the pass pair \{\emph{Op\-ti\-mize\-1q\-Gates\-Simple\-Com\-mutation}, \emph{Tem\-plate\-Op\-ti\-miza\-tion}\}, separated by the native gate set on the x-axis. The circuit depth is used as metric.}
    \label{fig:improve-order-ngs}
\end{figure}

In two of the seven cases (IDs 5 and 10) appending the first optimization pass to the sequence even leads to a visibly better result than the better permutation.
Fig.~\ref{fig:improve-order-bm} displays the results for the pass pair \{\emph{Op\-ti\-mize\-1q\-Gates\-Simple\-Com\-mutation}, \emph{Tem\-plate\-Op\-ti\-miza\-tion}\} regarding the circuit depth (the results for the gate count are analogous).
Again, the y-axis shows the values $R_{cd}(p_i), i\in\{0,1\}$ of the permutations, and the benchmarks are on the x-axis.
For each benchmark, there are four box plots for each sequence, from left to right: $p_0$, $p_1$, $p_{0+}$, and $p_{1+}$.
Using this pass pair, $p_1$ generally performs better than $p_0$.
When appending the first optimization pass to the worse sequence $p_0$, the resulting circuit gets better.
It depends on the benchmark how well $p_{0+}$ is in comparison to the circuit resulting from applying $p_0$: Looking at the benchmarks ae, qft, and randomcircuit, $p_{0+}$ improves the circuit in comparison to $p_0$, but not in comparison to $p_1$.
But when looking at the benchmarks dj, graphstate, and vqe\_two\_local, $p_{0+}$ returns the best result among the four tested sequences.
The sequence $p_{1+}$ in turn is only better than $p_1$, but not $p_{0+}$ when optimizing the graphstate benchmark; when optimizing randomcircuit and vqe\_two\_local, $p_{1+}$ performs worse than $p_1$.
Interestingly, the observation separating by the native gate set resembles the one with the benchmarks, but is even more specific: Fig.~\ref{fig:improve-order-ngs} presents the same results as Fig.~\ref{fig:improve-order-bm}, separated by the native gate set instead of the benchmark on the x-axis.
The order of passes $p_1$ is still generally better than $p_0$.
It shows that $p_{0+}$ is the overall best among the optimization sequences only when using the Quantinuum native gate set.
Sequence $p_{1+}$ performs worse than $p_1$ in the IQM and Quantinuum gate set and in the other native gate sets, appending the first optimization pass does not change the resulting circuit.
This emphasizes that the native gate set is relevant for the order of optimizations.

\begin{tcolorbox}[mytakeaway]
In some cases, it is possible to correct a less performing order of passes such that it performs at least as good as the best performing one.
The correction is not always successful and would lead to an overhead since there is then no change when appending the first optimization pass to the sequence.
\end{tcolorbox}

\subsection{Multi-Pass Sequences}\label{subsec:multipass_deps}

\subsubsection{Research Objectives}

Having established that pairwise interactions between optimization passes influence the performance of quantum circuit optimization, we extend our analysis to sequences containing more than two passes. 
The underlying hypothesis is that the identified pairwise interactions also affect longer pass sequences.
If an optimization pass pair returns different results depending on the application order, we call this a ``pairwise dependency''.
We systematically investigate all possible permutations of optimization pass sequences consisting of two to five passes.

\subsubsection{Experimental Setup}

For the experiments, all 16 optimization passes from Section~\ref{subsec:opt_passes} are taken into account.
%According to the symmetry of the binomial coefficient, for $k$ up to $\lfloor \frac{n}{2}\rfloor$, the number of combinations grows before decreasing again.
%In the case of $n=16$, the $k$ with the largest value is $\lfloor \frac{16}{2}\rfloor=7$ with  $c=\binom{16}{7}=6435$.
%Then the maximum number of optimization pass sequences to test for one $k$ in this case is $s=7!\cdot\binom{16}{7}=5040\cdot 11440=57,657,600$.
%Since running optimizations on circuits this often is not feasible, the largest $k$ that is used is $k=5$, i.e., $s=5!\cdot\binom{16}{5}=120\cdot 4368=524,160$.
We run all permutations for all combinations containing up to five optimization passes, therefore $s=5!\cdot\binom{16}{5}=
%120\cdot 4368=
524,160$.
Since this value grows superexponentially with the number of passes, conducting experiments using sequences with more than five passes is not feasible.
For time and memory reasons, the number of qubits for these experiments go from five to ten qubits instead of twelve qubits, but are still representative for the main findings.

\subsubsection{Experiment Results}

\begin{table}[]
    \settowidth\mylenA{Pairwise Dep. Count} 
    \setlength\mylenB{(\mylenA-2\tabcolsep)/2}
    \caption{Results from testing optimization pass sequences and their permutations with more than two passes. The table only shows the counts for sequences where the order of passes made a difference. For each number of passes, the minimum and maximum count of pairs with in the optimization pass sets and the minimum and maximum number of differing resulting circuits is shown.}
    \centering
    \small
    \begin{tabular}{|c|*{2}{wc{\mylenB}}|*{2}{wc{\mylenB}}|}
        \hline
        \multirow{2}{*}{\shortstack{Number of\\Passes}} & \multicolumn{2}{c|}{Pairwise Dep. Count} & \multicolumn{2}{c|}{Res. Circuits Count}\\
         & Min. & Max. & Min. & Max. \\\hline\hline
        2 & 1 & 1 & 2 & 2 \\\hline
        3 & 1 & 3 & 2 & 6 \\\hline
        4 & 1 & 6 & 2 & 15 \\\hline  
        5 & 1 & 9 & 2 & 32 \\\hline
    \end{tabular}
    \label{tab:multi-pass-deps}
\end{table}

Table~\ref{tab:multi-pass-deps} depicts the minimum and maximum number of pairwise dependencies in the sequence for two- to five-pass sequences in the second column.
Each row displays the results for the respective numbers of passes in one sequence.
The results reveal that pairwise relations are always present in an optimization sequence with at least two passes when the order has an impact on the resulting circuit. 
%This is always the case when there are exactly two passes which are order-sensitive.
The last column shows that the number of differing result circuits increases as the number of pairwise interactions grows, indicating multiple interactions between optimization passes. 

Considering the shortest best-performing optimization pass sequence for the specific native gate set, benchmark, and benchmark size combined, the results partly vary from the pairwise ones in Section~\ref{subsec:pairwise_deps}.
We find that \emph{Tem\-plate\-Op\-ti\-miza\-tion} in combination with both \emph{Op\-ti\-mize\-1q\-Gates\-Simple\-Com\-mutation} and \emph{Op\-ti\-mize\-1q\-Gates\-De\-com\-po\-si\-tion} is not at the beginning of the sequence, as IDs 5 and 10 of Table~\ref{tab:comb_deps} suggests, but between those two passes. 
We additionally find that ID 3 needs to be reversed in combination with ID 6 to output the best circuit, whereas ID 3 combined with ID 5 is just as previously suggested.
When combining ID 1 and ID 10, the latter one needs to be reversed to get an optimal result.

The experiments give more insight into which aspects are relevant to find the best optimization sequence:
With Rigetti's native gate set, the best sequence is independent of the benchmark and benchmark size and only depends on the native gate set.
With IBM Falcon and IonQ, the native gate set and the benchmark make a difference which optimization sequence performs the best, but the circuit size does not. 
Regarding the native gate sets IQM and Quantinuum, the native gate set, the circuit, and the circuit size affect which optimization sequence should be used to get the best result.
Solely the circuit or the circuit size do not influence the best optimization sequence, but solely the native gate set does.

\begin{tcolorbox}[mytakeaway]
Not only the order of individual pairs of optimizations affect the outcome, but the pairs can also influence each other.
This indicates multi-pass relations in addition to the pairwise ones.
\end{tcolorbox}

% \subsection{Initial Experiments with the t$|$ket$\rangle$ and Cirq Compiler}
% TODO?

\section{Discussion}\label{sec:dicussion}
%%%%%%%%%%%%%%%%%%%%%%%%%%%%%%%
%\RONE{The analysis of why particular pass interactions help or hurt remains somewhat limited. More mechanistic (quantum info. theoretic) explanation for representative cases would improve the paper's insight value and help readers transfer the lessons to other compilers. \\ -- Let's see what we can do.}
%\RTWO{The work would benefit from a deeper explanation of *why* certain pass pairs interact.  Currently, most conclusions are observational, which is ok.  But a discussion connecting the behavior of the passes to the structure of the transformations would strengthen the contribution. \\ -- Let's see what we can do.}
%%%%%%%%%%%%%%%%%%%%%%%%%%%%%%%

The results reveal that in 12 out of 120 considered combinations, there is a difference between the resulting optimized quantum circuits.
This implies that the order of optimization passes affects the quality of the resulting quantum circuit: 
Depending on the order of the optimizations, the quantum circuit has a smaller circuit depth and/or less gates and, hence, the pass sequence makes it less error-prone.

Correcting a suboptimal sequence by appending its first optimization pass is possible in 7 of 12 cases and in 2 of those cases, this results in a better optimized circuit.
%A non-optimal order can not always be corrected such that it optimizes the circuit as well as the best performing order of passes.
%When an order that is not best performing, i.e., it optimizes the circuit not as well as possible with the given optimization pass set, it can not always be corrected such that it optimizes the circuit as well as the best performing order of passes.
This implies that in particular cases and for specific sequences, it makes sense to repeat already applied passes to further optimize the circuit.
Repeating already applied passes also may worsen the outcome.
This demonstrates that using the shorter sequence is preferable, since appending another gate only produces an overhead and does not give any improvements.
Thus, initially choosing the best optimization order is important to get the best possible optimization.
Since reapplying the first optimization pass on the best performing two-pass sequence in some cases results in a better circuit regarding the considered metrics, it may make sense to use optimization sequences where passes are repeated after applying one or more other optimizations in-between.

% From the results from Sections~\ref{subsec:pairwise_deps} and~\ref{subsec:correct_passorder} we can also deduce that the native gate set plays a bigger role than the benchmarks when finding the right order of passes: The effect of the order of optimization passes varies for different native gate sets such that for some, the effect is significant and for some, there is only a light tendency.
% However, the tendency still persists in the sense that the better permutation is better in all cases, at least when not taking the correction into consideration.
% When taking the correction into consideration, it depends on the benchmark circuit and the native gate set whether appending the first optimization pass improves the resulting circuit in comparison to the shorter sequences.
% Hence, that to identify the best order of optimization passes, the native gate set of the target hardware has to be considered.
% This is because the native gate set determines which gates are included in the circuit, and with that the space of what can be optimized is therefore already given.

In our evaluations, sequences that contain more than two optimization passes include at least one pairwise dependency when finding that the order makes a difference.
An optimal order of passes may change when adding more optimization passes into the sequence, such that some pairwise dependencies might need to be reversed.
Thus, the identified pass pairs in Section~\ref{subsec:pairwise_deps} always have an impact on the optimized circuit, but the order might change depending on other optimization passes in the sequence.

While this work focuses on Qiskit's compiler, preliminary results from other compilers, such as t$|$ket$\rangle$ and Cirq, show similar trends. 
The analyses are not sufficiently detailed to be included as full results but suggests consistent conclusions.

%In the following, we propose a recommendation for the order of optimization passes derived from the results of this paper, and then compare this derived recommendation to the default order of Qiskit's compiler.

\subsection{Recommendation for Order of Optimization Passes}
%\RTWO{The recommendation for a "good" ordering should be presented more cautiously. Figure 6 shows that the proposed ordering does not clearly dominate Qiskit's default ordering.  Emphasizing that the recommendation is heuristic and context-dependent would make the conclusions more balanced. \\ -- Do that!}

The following recommendation of a good order of optimization passes\footnote{Note that this recommended order is based on the results and may change when considering other factors or passes.} is derived from the combined insights gained from Section~\ref{subsec:pairwise_deps} and %best performing order of the pairwise combinations that have a difference in their resulting circuits when applied in different order (Table~\ref{tab:comb_deps}) and the insights gained from
Section~\ref{subsec:multipass_deps}:
\begin{center}
    [\emph{Re\-move\-I\-den\-ti\-ty\-E\-qui\-va\-lent},
    \emph{Op\-ti\-mize\-1q\-Gates\-Sim\-ple\-Com\-mu\-ta\-tion},
    \emph{Tem\-plate\-Op\-ti\-miza\-tion},
    \emph{Com\-mu\-ta\-tive\-Can\-cel\-la\-tion},
    \emph{Op\-ti\-mize\-1q\-Gates\-De\-com\-po\-si\-tion},
    \emph{Com\-mu\-ta\-tive\-In\-verse\-Can\-cel\-la\-tion},
    \emph{Re\-move\-Di\-ag\-o\-nal\-Gates\-Be\-fore\-Mea\-sure},
    \emph{Op\-ti\-mize\-1q\-Gates}]
\end{center}
% \begin{enumerate}[label=(\roman*)]
%     \item \emph{Re\-move\-I\-den\-ti\-ty\-E\-qui\-va\-lent}
%     \item \emph{Op\-ti\-mize\-1q\-Gates\-Sim\-ple\-Com\-mu\-ta\-tion}
%     \item \emph{Tem\-plate\-Op\-ti\-miza\-tion}
%     \item \emph{Com\-mu\-ta\-tive\-Can\-cel\-la\-tion}
%     \item \emph{Op\-ti\-mize\-1q\-Gates\-De\-com\-po\-si\-tion}
%     \item \emph{Com\-mu\-ta\-tive\-In\-verse\-Can\-cel\-la\-tion}
%     \item \emph{Re\-move\-Di\-ag\-o\-nal\-Gates\-Be\-fore\-Mea\-sure}
%     \item \emph{Op\-ti\-mize\-1q\-Gates}
% \end{enumerate}
First apply \emph{Re\-move\-I\-den\-ti\-ty\-E\-qui\-va\-lent} to directly remove the gates that have negligible effects; those then do not influence the remaining optimizations.
Then \emph{Op\-ti\-mize\-1q\-Gates\-Sim\-ple\-Com\-mu\-ta\-tion} finds a suitable commutation of single-qubit gates through two-qubit gates to then conduct a simplification with these single-qubit gates and provides new space for pattern simplification. 
\emph{Tem\-plate\-Op\-ti\-miza\-tion} simplifies patterns and needs to be applied as soon as possible because patterns may disappear at a later point of time.
\emph{Com\-mu\-ta\-tive\-Can\-cel\-la\-tion} cancels self-adjoint gates using commutation relations, then 
\emph{Op\-ti\-mize\-1q\-Gates\-De\-com\-po\-si\-tion} simplifies chains of single-qubit gates before \emph{Com\-mu\-ta\-tive\-In\-verse\-Can\-cel\-la\-tion} cancels inverse gates using commutation relations.
\emph{Tem\-plate\-Op\-ti\-miza\-tion} comes between \emph{Op\-ti\-mize\-1q\-Gates\-Sim\-ple\-Com\-mu\-ta\-tion} and \emph{Op\-ti\-mize\-1q\-Gates\-De\-com\-po\-si\-tion} because it may find more patterns after a first simplification, and provides more space for another simplification afterwards.
As shown in Table~\ref{tab:comb_deps}, the differences between the permutations of the two cancellation passes and \emph{Op\-ti\-mize\-1q\-Gates\-De\-com\-po\-si\-tion} are very close to 0 with a tendency to the chosen order, so for those, it is not obvious which order is the best.
We assume that this is because the differences are this small because all passes act on the same kinds of gates and might in some cases interfere with each other. 
%For instance, if \emph{Op\-ti\-mize\-1q\-Gates\-De\-com\-po\-si\-tion} merges gates into two different blocks but the gates that may be canceled in these blocks might be now be incorporated in these blocks respectively. 
%On the other hand, the commutation of single-qubit gates might dissolve chains of single-qubit gates such that a simplification is not possible anymore that may have been possible before.
We choose the order of these three passes based on the derivations from Section~\ref{subsec:pairwise_deps} but note that the best order is not obvious for these.
%Then \emph{Op\-ti\-mize\-1q\-Gates\-Sim\-ple\-Com\-mu\-ta\-tion} finds a suitable commutation of single-qubit gates through two-qubit gates to then conduct a simplification with these single-qubit gates. 
%It needs to come after the simplification without commutations and cancellations because a more useful, and therefore better, commutation for another simplification is easier to find after already having simplified the circuit.
%Furthermore, the commutation can then not interfere with simplifications that would have been possible before commuting the gates.
Diagonal gates before measurements are removed with \emph{Re\-move\-Di\-ag\-o\-nal\-Gates\-Be\-fore\-Mea\-sure} after other removes, cancellations, and simplifications to not interfere with those and remove not needed diagonal gates before measurements.
Finally, \emph{Op\-ti\-mize\-1q\-Gates} merges all single-qubit gates together to one single gate. 
This is done in the end since none of the optimizations would have an effect on one single unitary gate.

From this recommended order, we derive a rough formalization on how to pick the order with given passes:
\begin{enumerate}[label=(\roman*)]
    \item Remove without constraints
    \item Commute and Simplify
    \item Pattern Matching
    \item Commute and Cancel or Simplify
    \item Commute and Simplify
    \item Remove with constraints
    \item Merge
\end{enumerate}
The optimization pass classes used here are partly derived from~\cite{quantum7010002,swierkowska24}. 
\emph{Remove} passes remove single gates without replacement. 
There are \emph{Remove} passes with and without constraints: Some require another gate type that influences whether the optimization is applied; these are the ones ``with constraints'', while the ones that do not need any other information than the existence of the gate to be removed are considered as the ones ``without constraints''.
\emph{Pattern Matching} passes replace gate patterns with an equivalent sequence.
\emph{Cancel} passes are similar to \emph{Remove} passes, only that they find two gates that cancel each other out instead of single gates.
Passes of the \emph{Simplify} class replace chains of gates with other equal gates and also include decompositions for the gates to be simplified.
\emph{Merge} passes merge several gates into one gate.
Since some optimizations include the commutation of gates before the optimization, the classes can be extended with \emph{Commute} as well.

Note that this formalization is derived from the 16 optimization passes considered in this study and may not fully capture the behavior of all possible passes. 
Furthermore, we found that several factors influence the best optimization sequence, so the recommended order does not always reflect the individual best permutation.
It provides a foundation for future generalization and can be extended as additional passes are incorporated.

\subsection{Comparison to Qiskit's Default Order of Passes}
%\RTWO{It would be interesting to compare the proposed ordering against automated pass-selection approaches or learning-based methods.  Such a comparison would help position the contribution relative to existing work. \\ -- Put this into future work because now this recommendation can still change because it is based only on one subset.}

% There is none since we are only changing the order, so the set of passes is still the same.
As mentioned in Section~\ref{subsec:sota-compilers}, Qiskit's compiler has a predefined default order of applying optimizations, depending on the optimization level.
Since optimization level 0 means no optimization, there are no optimizations at all.
In optimization level 1, the sequence is the following:
\begin{center}
    [\emph{InverseCancellation}, \emph{ContractIdleWiresInControlFlow}]
\end{center}
% \begin{enumerate}[label=(\roman*)]
%     \item \emph{InverseCancellation}
%     \item \emph{ContractIdleWiresInControlFlow}
% \end{enumerate}
The experiments show that these passes do affect each other and therefore, this order of optimizations is good as it is.
Optimization levels 2 and 3 use the following sequence:
\begin{center}
    [\emph{ElidePermutations}, \emph{Re\-move\-Di\-ag\-o\-nal\-Gates\-Be\-fore\-Mea\-sure}, \emph{Re\-move\-I\-den\-ti\-ty\-E\-qui\-va\-lent}, \emph{In\-verse\-Can\-cellation}, \emph{Con\-tract\-Idle\-Wires\-In\-Con\-trol\-Flow}, \emph{Com\-mu\-ta\-tive\-Can\-cel\-la\-tion}, \emph{Con\-soli\-date\-Blocks}, \emph{Split\-2Q\-Uni\-taries}]
\end{center}
% \begin{enumerate}[label=(\roman*)]
%     \item \emph{ElidePermutations}
%     \item \emph{RemoveDiagonalGatesBeforeMeasure}
%     \item \emph{RemoveIdentityEquivalent}
%     \item \emph{InverseCancellation}
%     \item \emph{ContractIdleWiresInControlFlow}
%     \item \emph{CommutativeCancellation}
%     \item \emph{ConsolidateBlocks}
%     \item \emph{Split2QUnitaries}
% \end{enumerate}
This does not match with the recommendation but the experiment results show that the passes used in Qiskit's optimization sequence do not have any effects on each other.
Therefore, the optimization result does not change when applying them in our recommended order [\emph{E\-lide\-Per\-mu\-ta\-tions}, \emph{Re\-move\-I\-den\-ti\-ty\-E\-qui\-va\-lent}, \emph{In\-verse\-Can\-cel\-la\-tion},  \emph{Con\-tract\-I\-dle\-Wires\-In\-Con\-trol\-Flow}, \emph{Com\-mu\-ta\-tive\-Can\-cel\-la\-tion}, \emph{Re\-move\-Di\-ag\-o\-nal\-Gates\-Be\-fore\-Mea\-sure}, \emph{Con\-so\-li\-date\-Blocks}, \emph{Split\-2Q\-U\-ni\-ta\-ries}].
The experiments in an isolated environment, i.e., only using the two optimization sequences respectively without any other kinds of compilation passes, show exactly this: Using Qiskit's order or our recommended order does not make any difference. 

\begin{figure}
    \centering
    \includegraphics[width=0.5\textwidth]{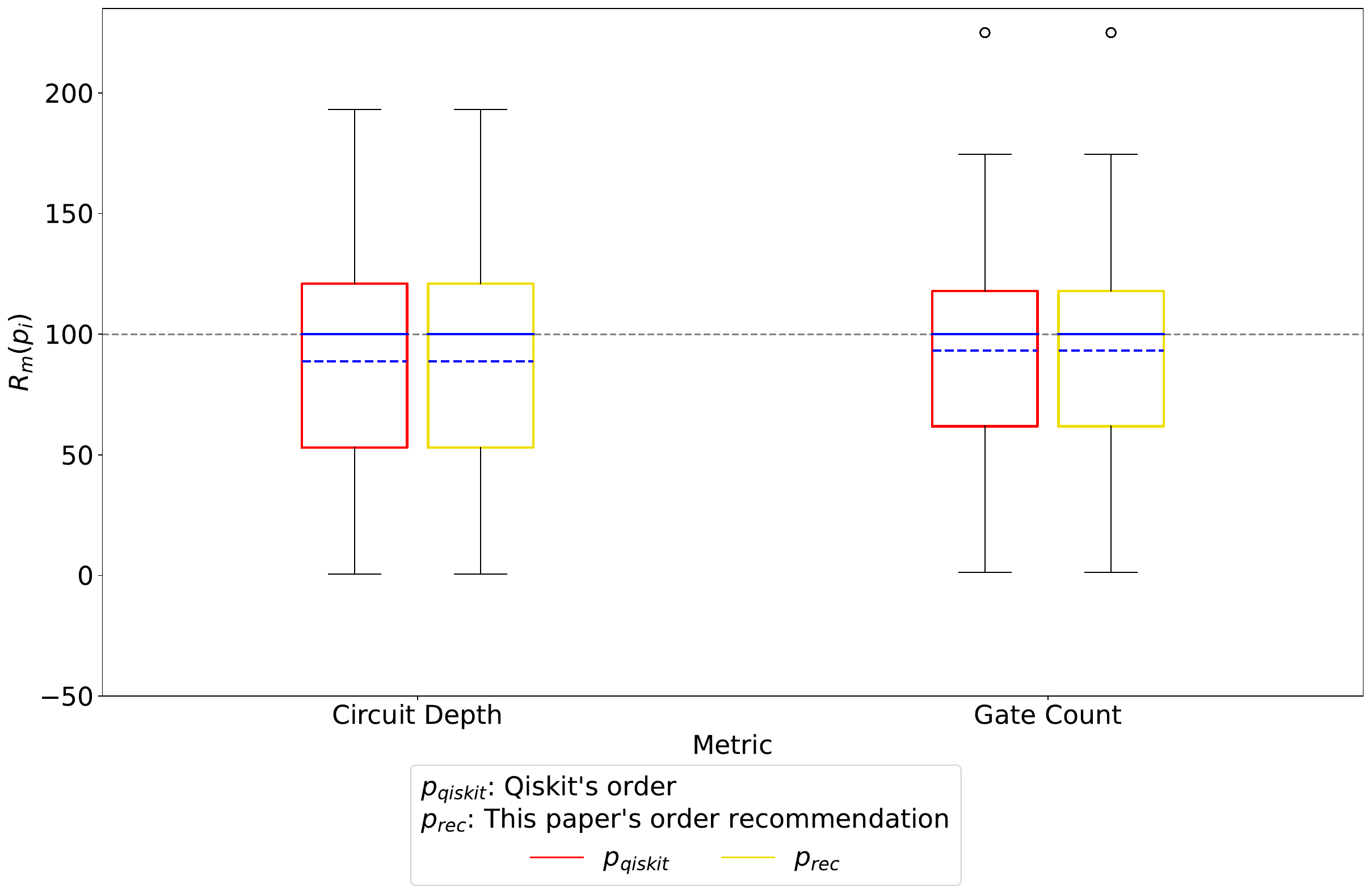}
    \caption{Box plots illustrating $R_m(p_i)$ for Qiskit's order $p_{qiskit}$ and this work's recommended order $p_{rec}$, subdivided by the used metrics in the x-axis.}
    \label{fig:comp-qiskit}
\end{figure}

We also compare the two orders without isolation by directly changing the order in Qiskit's compiler. 
It shows that in that case there is a difference in the resulting circuit when using these orders but it is not clear which order of optimization passes is the better one.
Fig.~\ref{fig:comp-qiskit} shows the results of the experiments using Qiskit's full compilation chain where we only change the order of optimization passes, with $R_m(p_i)$ ($p_{qiskit}$: Qiskit's order; $p_{rec}$: our recommended order) on the y-axis and the metrics on the x-axis.
For each metric, there are two box plots: The left one (red) is for Qiskit's order and the right one (yellow) is for our recommended order.
According to the box plots, there is no difference between the two permutations, since the minimum, maximum, and median for both are the same, and the mean only differs in the third decimal place.
In the median and mean, our recommended order is minimally better using the gate count as metrics.

The differences can only be observed in direct comparison: %Table~\ref{tab:comp-qiskit} shows how often Qiskit's order resulted in a better circuit (second column) regarding the circuit depth and the gate count respectively and the same for our recommended order (third column).
% \begin{table}[]
%     \caption{Numbers of cases where Qiskit or our recommended order respectively are better in the full compilation chain regarding the metrics circuit depth and gate count.}
%     \centering
%     \small
%     \begin{tabular}{|c|cc|}
%         \hline
%         Metric & Qiskit's Order & Our Order\\\hline\hline
%         Circuit Depth & 17 & 15 \\\hline
%         Gate Count & 5 & 9 \\\hline
%     \end{tabular}
%     \label{tab:comp-qiskit}
% \end{table}
In 17 cases, Qiskit's order is better than ours and in 15 cases our recommended order is better regarding the circuit depth, so Qiskit's order works better in more cases than ours for this metric.
Regarding the gate count, our recommended order is better than Qiskit's in 9 cases and Qiskit's is better in 5 cases, so our order works better in more cases than Qiskit's for this metric.
In conclusion, none of the two permutations is definitely better than the other one in general.

This shows that other compilation passes influence %the resulting optimized circuit, and therefore 
the best order of optimization passes.
The reason most probably is that a compilation chain includes other passes that transform the quantum circuit.
It is likely that these transformations affect the optimization and influence the optimization order.
\begin{tcolorbox}[mytakeaway]
We derive a recommended order of optimizaton passes from our experiment observations.
%The recommended order of optimization passes derived from our observations is the following: 
%[\emph{Re\-move\-I\-den\-ti\-ty\-E\-qui\-va\-lent}, \emph{Op\-ti\-mize\-1q\-Gates\-Sim\-ple\-Com\-mu\-ta\-tion}, \emph{Tem\-plate\-Op\-ti\-miza\-tion}, \emph{Com\-mu\-ta\-tive\-In\-verse\-Can\-cel\-la\-tion}, \emph{Com\-mu\-ta\-tive\-Can\-cel\-la\-tion}, \emph{Op\-ti\-mize\-1q\-Gates\-De\-com\-po\-si\-tion}, \emph{Op\-ti\-mize\-1q\-Gates}].
Compared to Qiskit's default order in the isolated case, i.e., when the pass sequence does not include other types of passes than optimization passes, there is no difference between the permutations. 
In the non-isolated case, there is a difference but no definite better order of optimizations. %, even though there is a minimal tendency to our recommended order.
%This means that most probably other compilation passes that are not optimization passes also influence the circuit optimization.
\end{tcolorbox}

\subsection{Factors Affecting the Optimal Order of Optimizations}

We identify some factors that affect the outcome of an optimization sequence and how well it works compared to its permutations.
While these factors are important, they do not represent an exhaustive list of all factors.

%From the results from Sections~\ref{subsec:pairwise_deps} and~\ref{subsec:correct_passorder}, t
The native gate set is in several cases relevant when finding the best order of passes: The effect of the order of optimization passes varies such that for some native gate sets, the effect is significant and for others, there is a light tendency or no effect at all.
The tendency still persists in the sense that the better permutation is better in all cases, at least when not considering correction.
The reason is that the native gate set determines which gates are included in the circuit, and with that the space of what can be optimized.
Regarding the given circuit, there are some cases where it is relevant, but it is not as apparent and not as strong as the relevance of the native gate set.

Considering correction, it depends on the circuit and the native gate set whether appending the first pass improves the resulting circuit compared to the shorter sequences.

Using more than two optimization passes, we find that the circuit size is an additional factor that is important to determine the best optimization sequence.

Comparing the full compilation chain using Qiskit's default order of passes and our recommended one, it becomes clear that earlier and/or later compilation passes influence how well an order of optimization passes performs.
%This means that the given circuit is important for the optimization sequence and needs to be taken into account, even if not as strongly as the native gate set.
This means that earlier and later transformations (which are not optimizations) need to be considered.

%Hence, to identify the best order of optimization passes, amongst others, the native gate set of the target hardware, the given circuit, and the later transformations must be regarded.

\begin{tcolorbox}[mytakeaway]
To identify the best order of optimization passes for a specific case, amongst others, the native gate set of the target hardware, the given circuit, its size, and earlier and later compiler passes must be considered.
\end{tcolorbox}

\section{Conclusion and Future Work}\label{sec:conclusion_fw}
%\RONE{Since the paper identifies strong ordering effects, it would be valuable to include a simple automated search or heuristic policy for discovering good pass sequences rather than leaving the contribution entirely descriptive. \\ -- Put this into future work}

In this paper, we present a thorough analysis on the impact of the order of optimization passes in quantum circuit compilation. 
Our systematic evaluation considers the metrics circuit depth and gate count, employing different optimization sequences within the Qiskit compilation environment.

Our results demonstrate that the order of optimization passes influences the resulting quantum circuit's quality. 
Specifically, out of the 120 analyzed pass combinations, 12 pairwise combinations show measurable differences based on the permutation of the order of passes. 
We observe preferences for certain orders of optimization passes, leading to the following recommended sequence: [\emph{Re\-move\-I\-den\-ti\-ty\-E\-qui\-va\-lent}, \emph{Op\-ti\-mize\-1q\-Gates\-Sim\-ple\-Com\-mu\-ta\-tion}, \emph{Tem\-plate\-Op\-ti\-miza\-tion}, \emph{Com\-mu\-ta\-tive\-Can\-cel\-la\-tion}, \emph{Op\-ti\-mize\-1q\-Gates\-De\-com\-po\-si\-tion}, \emph{Com\-mu\-ta\-tive\-In\-verse\-Can\-cel\-la\-tion}, \emph{Re\-move\-Di\-ag\-o\-nal\-Gates\-Be\-fore\-Mea\-sure}, \emph{Op\-ti\-mize\-1q\-Gates}]. 
The recommended order does not align with Qiskit's default sequence at optimization levels 2 or 3 but the results show that these specific passes do not have an impact on each other. 
%The optimization passes included in this sequence do not have any impact on each other, according to our pairwise experiments.
%Evaluations within an isolated environment (excluding additional compilation passes) validate our observations that these specific optimization passes do not have an impact on each other.
%In a non-isolated environment, the order of passes makes a difference, and therefore, previous and/or later compilation passes likely have an impact on the optimization sequence as well.
Evaluations within a non-isolated environment reveal that the order of passes makes a difference, implying that previous and/or later compilation passes likely have an impact on the optimization sequence as well.

We explore if less performing optimization sequences can be corrected by reapplying the first pass after the sequence. 
This approach succeeded in seven out of twelve sequences, and in two cases surpasses the originally best-performing sequence. 
In some cases, a repeated application of an optimization pass leads to a worse performance.
This underlines the necessity of initially choosing an optimal order of optimizations to avoid getting worse optimized circuits and to not create an overhead.

The analysis on optimization sequences with more than two passes shows that pairwise dependencies consistently emerge, suggesting their persistent impact. 
The shortest sequences resulting in the best-optimized circuit show that multiple optimizations affect each other and influence the best order of passes such that they partly differ from pairwise observations.

We identify that, amongst others, the following factors are important to determine the best optimization sequence for a given circuit: the target hardware's native gate set, the given circuit itself, its size, and earlier and later compiler passes.

In future work, the impact of other compilation passes on the optimization sequence should be assessed to clarify interactions within the compilation. 
In the process, additional metrics like fidelity-oriented ones and, if the specific hardware examined, hardware-related metrics may be considered.
With these insights, a more adaptive policy and deeper comparison to other proposed orderings would be a good next step.
%Additionally, comparative studies involving optimization sequences in other quantum compilers such as PennyLane~\cite{bergholm2022pennylaneautomaticdifferentiationhybrid}, Cirq~\cite{Cirq_Developers_2024}, tket~\cite{sivarajah_tket_2020}, or BQSKit~\cite{bqskit} are recommended. 
%This will clarify the interactions in the compilation workflow and further distinguish between target-agnostic and target-dependent optimizations.
Furthermore, investigating repeating optimization passes to find when a later repetition improves the resulting circuit would be interesting.

\section*{Acknowledgment}

The research is part of the Munich Quantum Valley (MQV), which is supported by the Bavarian state government with funds from the Hightech Agenda Bayern Plus. 
The work by A. E. is supported by BMW Group.
N. Q. and R. W. acknowledge funding from the European Research Council (ERC) under the European Union’s Horizon 2020 research and innovation program (grant agreement No. 101001318).

% \section*{References}

% Please number citations consecutively within brackets \cite{b1}. The 
% sentence punctuation follows the bracket \cite{b2}. Refer simply to the reference 
% number, as in \cite{b3}---do not use ``Ref. \cite{b3}'' or ``reference \cite{b3}'' except at 
% the beginning of a sentence: ``Reference \cite{b3} was the first $\ldots$''

% Number footnotes separately in superscripts. Place the actual footnote at 
% the bottom of the column in which it was cited. Do not put footnotes in the 
% abstract or reference list. Use letters for table footnotes.

% Unless there are six authors or more give all authors' names; do not use 
% ``et al.''. Papers that have not been published, even if they have been 
% submitted for publication, should be cited as ``unpublished'' \cite{b4}. Papers 
% that have been accepted for publication should be cited as ``in press'' \cite{b5}. 
% Capitalize only the first word in a paper title, except for proper nouns and 
% element symbols.

% For papers published in translation journals, please give the English 
% citation first, followed by the original foreign-language citation \cite{b6}.

\bibliographystyle{IEEEtran}
\bibliography{bibliography}

@inproceedings{grover96,
author = {Grover, Lov K.},
title = {A fast quantum mechanical algorithm for database search},
year = {1996},
isbn = {0897917855},
publisher = {Association for Computing Machinery},
address = {New York, NY, USA},
url = {https://doi.org/10.1145/237814.237866},
doi = {10.1145/237814.237866},
booktitle = {Proceedings of the Twenty-Eighth Annual ACM Symposium on Theory of Computing},
pages = {212–219},
numpages = {8},
location = {Philadelphia, Pennsylvania, USA},
series = {STOC '96}
}

@article{shor97,
author = {Shor, Peter W.},
title = {Polynomial-Time Algorithms for Prime Factorization and Discrete Logarithms on a Quantum Computer},
journal = {SIAM Journal on Computing},
volume = {26},
number = {5},
pages = {1484-1509},
year = {1997},
doi = {10.1137/S0097539795293172},
URL = { https://doi.org/10.1137/S0097539795293172},
eprint = { https://doi.org/10.1137/S0097539795293172}
}

@article{Chong2017,
  author       = {Frederic T. Chong and Diana Franklin and Margaret Martonosi},
  title        = {Programming languages and compiler design for realistic quantum hardware},
  journal      = {Nature},
  volume       = {549},
  number       = {7671},
  pages        = {180--187},
  year         = {2017},
  publisher    = {Nature Publishing Group},
  doi          = {10.1038/nature23459},
  url          = {https://www.nature.com/articles/nature23459}
}

@manual{gcc,
  title        = {{GCC}, the {GNU} Compiler Collection},
  author       = {{GNU Project}},
  organization = {Free Software Foundation},
  year         = {2024},
  note         = {[Online]. Available: \url{https://gcc.gnu.org/}},
}

@INPROCEEDINGS{LLVM,
  author={Lattner, C. and Adve, V.},
  booktitle={International Symposium on Code Generation and Optimization, 2004. CGO 2004.}, 
  title={LLVM: a compilation framework for lifelong program analysis \& transformation}, 
  year={2004},
  volume={},
  number={},
  pages={75-86},
  doi={10.1109/CGO.2004.1281665}}

@misc{BridgeTheGap,
author = {Elsharkawy, Amr and Guo, Xiaorang and Schulz, Martin},
title = "Bridge the Gap Between HPC Systems and Various Quantum Platforms: A Unified Quantum Platform",
year = 2025,
doi = "10.18420/se2025-ws-16",
howpublished = "Software Engineering 2025 – Companion Proceedings",
publisher = "Gesellschaft für Informatik, Bonn",
issn = "2944-7682",
eissn = "2944-7682",
}

@INPROCEEDINGS{UQP,
  author={Elsharkawy, Amr and Guo, Xiaorang and Schulz, Martin},
  booktitle={2024 IEEE International Conference on Quantum Computing and Engineering (QCE)}, 
  title={Integration of Quantum Accelerators into HPC: Toward a Unified Quantum Platform}, 
  year={2024},
  volume={01},
  number={},
  pages={774-783},
  doi={10.1109/QCE60285.2024.00097}}

@misc{HPCQCReview,
      title={Integration of Quantum Accelerators with High Performance Computing -- A Review of Quantum Programming Tools}, 
      author={Amr Elsharkawy and Xiao-Ting Michelle To and Philipp Seitz and Yanbin Chen and Yannick Stade and Manuel Geiger and Qunsheng Huang and Xiaorang Guo and Muhammad Arslan Ansari and Christian B. Mendl and Dieter Kranzlmüller and Martin Schulz},
      year={2023},
      eprint={2309.06167},
      archivePrefix={arXiv},
      primaryClass={cs.ET},
      url={https://arxiv.org/abs/2309.06167}, 
}

@INPROCEEDINGS{Seitz_HPCQC,
  author={Seitz, Philipp and Elsharkawy, Amr and To, Xiao-Ting Michelle and Schulz, Martin},
  booktitle={2023 IEEE International Conference on Quantum Computing and Engineering (QCE)}, 
  title={Toward a Unified Hybrid HPCQC Toolchain}, 
  year={2023},
  volume={02},
  pages={96-102},
  doi={10.1109/QCE57702.2023.10191},
  url={https://ieeexplore.ieee.org/document/10313648}
  }

@INPROCEEDINGS{HPCQCchallenges,
  author={Elsharkawy, Amr and To, Xiao-Ting Michelle and Seitz, Philipp and Chen, Yanbin and Stade, Yannick and Geiger, Manuel and Huang, Qunsheng and Guo, Xiaorang and Ansari, Muhammad Arslan and Ruefenacht, Martin and Schulz, Laura and Karlsson, Sven and Mendl, Christian B. and Kranzlmüller, Dieter and Schulz, Martin},
  booktitle={2023 IEEE International Conference on Quantum Computing and Engineering (QCE)}, 
  title={Challenges in HPCQC Integration}, 
  year={2023},
  volume={02},
  pages={405-406},
  doi={10.1109/QCE57702.2023.10304},
  url={https://ieeexplore.ieee.org/document/10313875}
}

@misc{qiskit,
	title        = {Qiskit: An open-source framework for quantum computing},
	author       = {Aleksandrowicz, Gadi and Alexander, Thomas and Barkoutsos, Panagiotis and Bello, Luciano and Ben-Haim, Yael and Bucher, David and Cabrera-Hern{\'a}ndez, F. Jose and Carballo-Franquis, Jorge and Chen, Adrian and Chen, Chun-Fu and others},
	year         = 2019,
	url          = {https://qiskit.org/}
}

@article{sivarajah_tket_2020,
	title        = {t{\textbar}ket⟩ : {A} {Retargetable} {Compiler} for {NISQ} {Devices}},
	shorttitle   = {t\${\textbar}\$ket\${\textbackslash}rangle\$},
	author       = {Sivarajah, Seyon and Dilkes, Silas and Cowtan, Alexander and Simmons, Will and Edgington, Alec and Duncan, Ross},
	year         = 2020,
	journal      = {Quantum Science and Technology},
	volume       = 6,
	number       = 1,
	pages        = {014003},
	doi          = {10.1088/2058-9565/ab8e92},
	issn         = {2058-9565},
	url          = {http://arxiv.org/abs/2003.10611},
}

@misc{bqskit,
  author       = {Younis, Ed and Iancu, Costin C and Lavrijsen, Wim and Davis, Marc and Smith, Ethan and USDOE},
  title        = {Berkeley Quantum Synthesis Toolkit (BQSKit) v1},
  doi          = {10.11578/dc.20210603.2},
  url          = {https://www.osti.gov/biblio/1785933},
  place        = {United States},
  year         = {2021},
  month        = {04}}

@software{cudaq,
  author       = {The CUDA-Q development team},
  title        = {CUDA-Q},
  month        = dec,
  year         = 2024,
  publisher    = {Zenodo},
  version      = {0.9.1},
  doi          = {10.5281/zenodo.14503457},
  url          = {https://doi.org/10.5281/zenodo.14503457},
  swhid        = {swh:1:dir:52dd2a3417ddb5a1002c79c04e3acde72a691cc2
                   ;origin=https://doi.org/10.5281/zenodo.8092233;vis
                   it=swh:1:snp:063dce7f790283c81583b91e278f5734a9099
                   927;anchor=swh:1:rel:30fd17b958fa9c4698b596b4a82d1
                   d4d02672343;path=NVIDIA-cuda-quantum-a9d4130
                  },
}

@INPROCEEDINGS{swierkowska24,
  author={Świerkowska, Aleksandra and Echavarria, Jorge and Schulz, Laura and Schulz, Martin},
  booktitle={2024 IEEE International Conference on Quantum Computing and Engineering (QCE)}, 
  title={Achieving Pareto-Optimality in Quantum Circuit Compilation via a Multi-Objective Heuristic Optimization Approach}, 
  year={2024},
  volume={02},
  number={},
  pages={306-310},
  doi={10.1109/QCE60285.2024.10297}}

@article{staudacher2023,
author = {Staudacher, Korbinian and Guggemos, Tobias and Grundner-Culemann, Sophia and Gehrke, Wolfgang},
year = {2023},
month = {11},
pages = {29-45},
title = {Reducing 2-QuBit Gate Count for ZX-Calculus based Quantum Circuit Optimization},
volume = {394},
journal = {Electronic Proceedings in Theoretical Computer Science},
doi = {10.4204/EPTCS.394.3}
}

@article{quetschlich2023mqtbench,
  title={{{MQT Bench}}: Benchmarking Software and Design Automation Tools for Quantum Computing},
  shorttitle = {{MQT Bench}},
  journal = {{Quantum}},
  author={Quetschlich, Nils and Burgholzer, Lukas and Wille, Robert},
  year={2023},
  note={{{MQT Bench}} is available at \url{https://www.cda.cit.tum.de/mqtbench/}},
}

@article{nam2018automated,
  title={Automated optimization of large quantum circuits with continuous parameters},
  author={Nam, Yunseong and Ross, Neil J and Su, Yuan and Childs, Andrew M and Maslov, Dmitri},
  journal={npj Quantum Information},
  volume={4},
  number={1},
  pages={23},
  year={2018},
  publisher={Nature Publishing Group UK London}
}

@ARTICLE{amy2014,
  author={Amy, Matthew and Maslov, Dmitri and Mosca, Michele},
  journal={IEEE Transactions on Computer-Aided Design of Integrated Circuits and Systems}, 
  title={Polynomial-Time T-Depth Optimization of Clifford+T Circuits Via Matroid Partitioning}, 
  year={2014},
  volume={33},
  number={10},
  pages={1476-1489},
  doi={10.1109/TCAD.2014.2341953}}

@misc{bergholm2022pennylaneautomaticdifferentiationhybrid,
      title={PennyLane: Automatic differentiation of hybrid quantum-classical computations}, 
      author={Ville Bergholm and Josh Izaac and Maria Schuld and Christian Gogolin and Shahnawaz Ahmed and Vishnu Ajith and M. Sohaib Alam and Guillermo Alonso-Linaje and B. AkashNarayanan and Ali Asadi and Juan Miguel Arrazola and Utkarsh Azad and Sam Banning and Carsten Blank and Thomas R Bromley and Benjamin A. Cordier and Jack Ceroni and Alain Delgado and Olivia Di Matteo and Amintor Dusko and Tanya Garg and Diego Guala and Anthony Hayes and Ryan Hill and Aroosa Ijaz and Theodor Isacsson and David Ittah and Soran Jahangiri and Prateek Jain and Edward Jiang and Ankit Khandelwal and Korbinian Kottmann and Robert A. Lang and Christina Lee and Thomas Loke and Angus Lowe and Keri McKiernan and Johannes Jakob Meyer and J. A. Montañez-Barrera and Romain Moyard and Zeyue Niu and Lee James O'Riordan and Steven Oud and Ashish Panigrahi and Chae-Yeun Park and Daniel Polatajko and Nicolás Quesada and Chase Roberts and Nahum Sá and Isidor Schoch and Borun Shi and Shuli Shu and Sukin Sim and Arshpreet Singh and Ingrid Strandberg and Jay Soni and Antal Száva and Slimane Thabet and Rodrigo A. Vargas-Hernández and Trevor Vincent and Nicola Vitucci and Maurice Weber and David Wierichs and Roeland Wiersema and Moritz Willmann and Vincent Wong and Shaoming Zhang and Nathan Killoran},
      year={2022},
      eprint={1811.04968},
      archivePrefix={arXiv},
      primaryClass={quant-ph},
      url={https://arxiv.org/abs/1811.04968}, 
}

@book{Cirq_Developers_2024, title={Cirq}, url={https://zenodo.org/doi/10.5281/zenodo.4062499}, DOI={10.5281/ZENODO.4062499}, abstractNote={Cirq is a Python library for writing, manipulating, and optimizing quantum circuits and running them against quantum computers and simulators.}, publisher={Zenodo}, author={Cirq Developers}, year={2024}, month=may }

@INPROCEEDINGS{kim2023cudaq,
  author={Kim, Jin-Sung and McCaskey, Alex and Heim, Bettina and Modani, Manish and Stanwyck, Sam and Costa, Timothy},
  booktitle={2023 60th ACM/IEEE Design Automation Conference (DAC)}, 
  title={CUDA Quantum: The Platform for Integrated Quantum-Classical Computing}, 
  year={2023},
  volume={},
  number={},
  pages={1-4},
  doi={10.1109/DAC56929.2023.10247886}}

@InProceedings{deb2000,
author="Deb, Kalyanmoy
and Agrawal, Samir
and Pratap, Amrit
and Meyarivan, T.",
editor="Schoenauer, Marc
and Deb, Kalyanmoy
and Rudolph, G{\"u}nther
and Yao, Xin
and Lutton, Evelyne
and Merelo, Juan Julian
and Schwefel, Hans-Paul",
title="A Fast Elitist Non-dominated Sorting Genetic Algorithm for Multi-objective Optimization: NSGA-II",
booktitle="Parallel Problem Solving from Nature PPSN VI",
year="2000",
publisher="Springer Berlin Heidelberg",
address="Berlin, Heidelberg",
pages="849--858",
isbn="978-3-540-45356-7"
}

@article{Iten_2022,
   title={Exact and Practical Pattern Matching for Quantum Circuit Optimization},
   volume={3},
   ISSN={2643-6817},
   url={http://dx.doi.org/10.1145/3498325},
   DOI={10.1145/3498325},
   number={1},
   journal={ACM Transactions on Quantum Computing},
   publisher={Association for Computing Machinery (ACM)},
   author={Iten, Raban and Moyard, Romain and Metger, Tony and Sutter, David and Woerner, Stefan},
   year={2022},
   month=jan, pages={1–41} }

@Article{quantum7010002,
AUTHOR = {Karuppasamy, Krishnageetha and Puram, Varun and Johnson, Stevens and Thomas, Johnson P.},
TITLE = {A Comprehensive Review of Quantum Circuit Optimization: Current Trends and Future Directions},
JOURNAL = {Quantum Reports},
VOLUME = {7},
YEAR = {2025},
NUMBER = {1},
ARTICLE-NUMBER = {2},
URL = {https://www.mdpi.com/2624-960X/7/1/2},
ISSN = {2624-960X},
DOI = {10.3390/quantum7010002}
}

@article{rahman2015,
author = {Rahman, Md. Mazder and Dueck, Gerhard W. and Horton, Joseph D.},
title = {An Algorithm for Quantum Template Matching},
year = {2015},
issue_date = {December 2014},
publisher = {Association for Computing Machinery},
address = {New York, NY, USA},
volume = {11},
number = {3},
issn = {1550-4832},
url = {https://doi.org/10.1145/2629537},
doi = {10.1145/2629537},
journal = {J. Emerg. Technol. Comput. Syst.},
month = dec,
articleno = {31},
numpages = {20}
}

@article{PhysRevA.102.022406,
  title = {Reducing the number of non-Clifford gates in quantum circuits},
  author = {Kissinger, Aleks and van de Wetering, John},
  journal = {Phys. Rev. A},
  volume = {102},
  issue = {2},
  pages = {022406},
  numpages = {10},
  year = {2020},
  month = {Aug},
  publisher = {American Physical Society},
  doi = {10.1103/PhysRevA.102.022406},
  url = {https://link.aps.org/doi/10.1103/PhysRevA.102.022406}
}

@inproceedings{touati2006,
author = {Touati, Sid-Ahmed-Ali and Barthou, Denis},
title = {On the decidability of phase ordering problem in optimizing compilation},
year = {2006},
isbn = {1595933026},
publisher = {Association for Computing Machinery},
address = {New York, NY, USA},
url = {https://doi.org/10.1145/1128022.1128042},
doi = {10.1145/1128022.1128042},
booktitle = {Proceedings of the 3rd Conference on Computing Frontiers},
pages = {147–156},
numpages = {10},
location = {Ischia, Italy},
series = {CF '06}
}

@article{Almakki_Ali_Izzeldin_Huang_Cummins, title={Autophase V2: Towards Function Level Phase Ordering Optimization}, abstractNote={Compilers are equipped with optimization passes that can be applied to improve the quality of a program. The selection and ordering of these passes is a classic NP-hard problem known as the phase ordering problem. Traditionally, compilers use expert-picked sequences to optimize for performance (e.g. O2, -O3), or code size (e.g. -Os, -Oz). However, not all programs respond positively to all optimizations, and prior work has shown that these expert-picked phase orders can be outperformed by tailoring the phase order decisions to individual programs.}, author={Almakki, Mohammed and Ali, Ameer Haj and Izzeldin, Ayman and Huang, Qijing and Cummins, Chris}, language={en} }

@ARTICLE{quetschlich2024mqtpredictor,
    AUTHOR      = {N. Quetschlich and L. Burgholzer and R. Wille},
    TITLE       = {{MQT Predictor: Automatic Device Selection with Device-Specific Circuit Compilation for Quantum Computing}},
    YEAR        = {2025},
    JOURNAL     = {ACM Transactions on Quantum Computing (TQC)},
    DOI         = {10.1145/3673241},
    EPRINT      = {2310.06889},
    EPRINTTYPE  = {arxiv},
}

@misc{dangwal2025cliffordassistedoptimalpass,
      title={Clifford Assisted Optimal Pass Selection for Quantum Transpilation}, 
      author={Siddharth Dangwal and Gokul Subramanian Ravi and Lennart Maximilian Seifert and Poulami Das and James Sud and Frederic T. Chong},
      year={2025},
      eprint={2306.15020},
      archivePrefix={arXiv},
      primaryClass={quant-ph},
      url={https://arxiv.org/abs/2306.15020}, 
}

@INPROCEEDINGS{peham2023clifford,
    TUMFIS = {a319a9d2b38e46068809018be80c0408},
    MEDIATUM = {a0dADdedzhM3qYovHXlqc oYmdZidoSij6WlixJO1MT},
	AUTHOR    = {T. Peham and N. Brandl and R. Kueng and R. Wille and L. Burgholzer},
	TITLE     = {{Depth-Optimal Synthesis of Clifford Circuits with SAT Solvers}},
	BOOKTITLE = qce,
	YEAR      = {2023},
	ACKS      = {erc,mqv,quast, scch},
	DOMAIN    = {quantum, quantum-synth}
}

@misc{huang2020autophasejugglinghlsphase,
      title={AutoPhase: Juggling HLS Phase Orderings in Random Forests with Deep Reinforcement Learning}, 
      author={Qijing Huang and Ameer Haj-Ali and William Moses and John Xiang and Ion Stoica and Krste Asanovic and John Wawrzynek},
      year={2020},
      eprint={2003.00671},
      archivePrefix={arXiv},
      primaryClass={cs.DC},
      url={https://arxiv.org/abs/2003.00671}, 
}

\end{document}